%% file: X-ray_dm_rocket.tex
\documentclass[numberedappendix,iop,apj,appendixfloats,twocolappendix]{emulateapj}

\usepackage{graphicx}
\usepackage{color}
\usepackage{amsmath}
\usepackage{hhline}
\usepackage{subfigure}
\usepackage{threeparttable}
\slugcomment{Draft Version \today}
\shorttitle{keV Dark Matter with Sounding Rockets}
\shortauthors{Figueroa-Feliciano et al.}

\newcommand{\ltsima}{$\; \buildrel < \over \sim \;$}
\newcommand{\simlt}{\lower.5ex\hbox{\ltsima}}

\newcommand{\Suzaku}{\emph{Suzaku}}
\newcommand{\XMM}{\emph{XMM-Newton}}
\newcommand{\Chandra}{\emph{Chandra}}
\newcommand{\ROSAT}{\emph{ROSAT}}
\newcommand{\Ofov}{$\Omega_{\mathrm{FOV}}$}
\newcommand{\Aeff}{$A_{\mathrm{eff}}$}

\begin{document}

\title{Searching for keV Sterile Neutrino Dark Matter with X-ray Microcalorimeter Sounding Rockets}

\author{E.~Figueroa-Feliciano,\altaffilmark{1}
A.~J.~Anderson,\altaffilmark{1}
D.~Castro, \altaffilmark{1}
D.~C.~Goldfinger,\altaffilmark{1}
J.~Rutherford\altaffilmark{1}\\ \vspace{5mm}
M.~E.~Eckart,\altaffilmark{2}
R.~L.~Kelley,\altaffilmark{2}
C.~A.~Kilbourne,\altaffilmark{2}
D.~McCammon,\altaffilmark{3}
K.~Morgan,\altaffilmark{3}
F.~S.~Porter,\altaffilmark{2}
A.~E.~Szymkowiak\altaffilmark{4}\\
{(XQC Collaboration)}
}
\altaffiltext{1}{Department of Physics and Kavli Institute for Astrophysics and Space Research, Massachusetts Institute of Technology, Cambridge, MA 02139, USA}
\altaffiltext{2}{NASA Goddard Space Flight Center, Greenbelt, MD 20771, USA}
\altaffiltext{3}{Department of Physics, University of Wisconsin, Madison, WI 53706, USA}
\altaffiltext{4}{Department of Physics, Yale University, New Haven, CT 06511, USA}

\email{enectali@mit.edu}
\begin{abstract}

High-resolution X-ray spectrometers onboard suborbital sounding rockets can search for dark matter candidates that produce X-ray lines, such as decaying keV-scale sterile neutrinos. Even with exposure times and effective areas far smaller than \XMM\ and \Chandra\ observations, high-resolution, wide field-of-view observations with sounding rockets have competitive sensitivity to decaying sterile neutrinos. We analyze a subset of the 2011 observation by the X-ray Quantum Calorimeter instrument centered on Galactic coordinates $l = 165^\circ, b = -5^\circ$ with an effective exposure of 106~seconds, obtaining a limit on the sterile neutrino mixing angle of $\sin^{2}2\theta < 7.2 \times 10^{-10}$ at 95\% CL for a 7~keV neutrino. Better sensitivity at the level of $\sin^{2}2\theta\sim 2.1 \times 10^{-11}$ at 95\% CL for a 7~keV neutrino is achievable with future 300-second observations of the galactic center by the Micro-X instrument, providing a definitive test of the sterile neutrino interpretation of the reported 3.56~keV excess from galaxy clusters. 
\end{abstract}


\input{intro.tex}
\input{XQClimit.tex}

\input{backgrounds.tex}

\input{Micro-X-sensitivity.tex}

\input{conclusion.tex}
\clearpage
\input{appendixXQC.tex}

\nocite{*}
\bibliographystyle{apj}
\bibliography{X-ray_DM_Rocket.bib}


\end{document}

%% file: intro.tex
\section{Introduction}
\label{sec:intro}
A variety of dark matter models predict photon production via dark matter decay, annihilation, or de-excitation. Some of these models predict mono-energetic photons with energies in the 1-100~keV range, prompting recent searches for lines in existing X-ray data from the \XMM\ and \Chandra\ observatories. Due to the well-understood atomic physics in this energy range and the ability to check the morphology of a potential signal against expectations from galactic dark matter halos, X-ray lines could provide unambiguous evidence for some models of astrophysical dark matter.


There has been heightened interest in dark matter searches in the X-ray band following claims of an unidentified X-ray line seen in both galaxy and galaxy cluster observations. \citet{Bulbul2014} analyzed \XMM\ observations of 73 stacked galaxy clusters and found an excess line with energy around 3.56 keV. The line is present at the $> 3\sigma$ level in three separate subsamples of data from both the MOS and PN instruments, and they also detect it in \Chandra\ observations of the Perseus cluster. \citet{Boyarsky2014} reported a $>3\sigma$ excess around 3.53~keV in their spectral fits of \XMM\ observations of the Perseus cluster and the Andromeda galaxy. In both cases, the width of the measured excess is determined by the \XMM\ and \Chandra\ instrument response.  Analysis of \XMM\ observations of the Milky Way Galactic Center (MW GC) by \citet{Boyarsky2014a} finds a formal $5.7\sigma$ excess at the expected energy. However, the complexity of the GC makes modeling the background difficult, and because of the instrumental resolution of the observation they cannot rule out the possibility of the excess coming from K XVIII emission.

A vigorous search has ensued, with various reports of non-detections: \citet{RiemerSorensen:2014us} in the GC, \citet{Jeltema:2014wr}  in the GC and M31, \citet{Anderson:2014wc} in galaxies and galaxy groups, and \citet{Malyshev:2014vf} in dwarf spheroidal galaxies. There has been some debate as to how to best fit the continuum and of what the allowed flux of astrophysical lines (primarily from K, Cl, and Ar) in the pertinent energy range should be \citep{Bulbul:2014wj,Jeltema:2014wr}. 

\citet{Urban:2014va} reproduce the line in Perseus using \Suzaku\ but find its spatial distribution in tension with expectations for decaying dark matter, and further they do not detect the expected scaled emission in either the Coma, Virgo or Ophiuchus clusters. \citet{Carlson:2014vl} performed a morphological study of the continuum-subtracted excess emission at 3.5~keV in the GC and Perseus. They find the GC excess spatial distribution incompatible with the expected DM distribution, and strongly correlated with the morphology of atomic lines from Ar and Ca with energies between 3--4~keV. The Perseus emission is correlated most strongly with the cool core emission, confirming the tension presented in the \Suzaku\ Perseus observation.

This set of observations demonstrates the challenges in searching for X-ray lines with the current observatories. New instruments are needed to improve the sensitivity of searches for X-ray line emission from dark matter. \citet{Boyarsky2006a} studied the optimal characteristics of a mission dedicated to searches of diffuse line emission in the X-ray regime. They pointed out that the main determinants of instrument sensitivity are \emph{grasp}~$\equiv$ \Aeff \Ofov\ (effective area $\times$ field of view, also referred to as \emph{\'etendue}) and \emph{energy resolution} $\Delta E/E$. As a ``prototype'' demonstration, \citet{Boyarsky2006a} calculated the limits on the sterile neutrino mixing angle $\sin^{2}2\theta$ for data from the third flight of the X-ray Quantum Calorimeter (XQC) sounding rocket payload \citep{McCammon:2002p160} for sterile neutrino masses between 0.4--2~keV. 

Existing X-ray telescopes tend to have comparatively small fields of view (e.g. a few tens of arcminutes for instruments on \XMM~and \Chandra) with insufficient energy resolution to resolve closely spaced weak spectral lines (e.g. $\sim 100$~eV FWHM at 2~keV for the EPIC camera on \XMM). Because of our location within the dark matter halo of the MW, the sterile neutrino decay is an \emph{all-sky signal}, so sensitivity can be improved by increasing the FOV. Discrimination of a signal against atomic lines can also be significantly improved using the superior energy resolution available with X-ray microcalorimeters. This combination of large-FOV with high spectral resolution is achieved in existing microcalorimeter payloads on sounding rockets. Although the exposure from a typical sounding rocket flight is less than 300~s, the sensitivity of these short observations can be competitive with deep \XMM\ observations of the GC. The upcoming SXS microcalorimeter instrument onboard ASTRO-H \citep{Takahashi2014} will have excellent $<7$~eV resolution, but its narrow 3'$\times$3' FOV limits its sensitivity to the all-sky signal expected from sterile neutrino decay in the MW.  Wide-FOV sounding rocket observations are therefore complementary to the deep ($\sim$1~Msec) observations of the cores of galaxy clusters, galaxies, and dwarf spheroidals that ASTRO-H will perform \citep{Kitayama2014}.

In this paper we set limits on decaying sterile neutrino dark matter using a new dataset from the XQC sounding rocket in order to demonstrate the reach and analysis of large-FOV observations, and we discuss the sensitivity and optimization of future observations using new instruments, such as the Micro-X detector. Section~\ref{sec:motivation} discusses the sterile neutrino signal and estimates the signal and background of large-FOV observations for a putative signal. In Section~\ref{sec:XQCdata} we describe the XQC instrument, present an analysis of data taken during the 5th flight of XQC, and place limits on the sterile neutrino mixing angle $\sin^{2}2\theta$ for sterile neutrino masses of 4--10~keV. Section~\ref{sec:future} estimates the sensitivity of observations with the future Micro-X payload by constructing a detailed background model based on ASCA, ROSAT, and Suzaku observations and analyzing mock data sets. Section~\ref{sec:sterilelimit} discusses the implications of our flux limits on sterile neutrino dark matter.

\section{keV Dark Matter with Rockets}
\label{sec:motivation}

\subsection{Dark Matter Interpretations of X-ray Lines}
\label{sec:theory}
Well-motivated dark matter models can produce X-ray lines through decay, de-excitation, or annihilation. Perhaps the best-known scenario is that of keV-mass sterile neutrinos \citep{Abazajian2001,Asaka2005,Boyarsky2006}, although a large number of models have been proposed following the observations of the 3.56~keV line. Such models include axions, axinos, exciting dark matter, gravitinos, moduli, and WIMPs, among others \citep[see discussion in][]{Jeltema:2014wr}.  

Sterile neutrinos and other models that produce photons by particle decay predict a flux per volume element that scales as the dark matter density ($\rho$). Other models, such as eXciting dark matter \citep{Finkbeiner2007,Finkbeiner2014,Berlin2015}, that require two dark matter particles to interact predict a line flux per volume element that scales as the dark matter density squared ($\rho^2$). More complex scenarios, involving eXciting dark matter with a primordial population in an excited state \citep{Finkbeiner2014}, can furthermore produce fluxes that scale as $\rho^{\alpha}$ with $1<\alpha<2$. A non-linear scaling implies a much smaller flux from lower density systems like dwarf spheroidal galaxies and could ease the tension with non-observations of the 3.5~keV line in these systems \citep{Malyshev:2014vf}. In this paper we will use sterile neutrinos as our benchmark model, although we present our results as a line flux limit from a particular target, which can be translated into a constraint or signal in any of these models. 

Sterile neutrinos with masses in the $\sim$1-100~keV range may contribute to the dark matter relic density if they are produced in the early universe. Two well-studied production mechanisms are non-resonant oscillation of active neutrinos \citep{Dodelson1994} and resonant oscillation via the MSW effect \citep{Shi1999}. The existence of sterile neutrinos is additionally motivated by neutrino oscillation data, which could be explained by adding sterile right-handed neutrinos to the standard model, such as in the $\nu$MSM scenario \citep{Asaka2005a}. Although they must possess cosmological lifetimes in order to contribute to the dark matter relic density, sterile neutrinos may decay to a photon and active neutrino via a loop-suppressed process mediated by oscillation between the active and sterile states. The rate for this process is given by \citep{Pal1982}
\begin{eqnarray}
\Gamma &=& \frac{9\alpha G_F^2 m_{s}^5 \sin^2 2\theta }{1024\pi^4}\\
 &=& (1.38 \times 10^{-29} \textrm{ s}^{-1}) \left( \frac{\sin^2 2\theta}{10^{-7}} \right) \left( \frac{m_s}{1\textrm{ keV}} \right)^5,\label{eqn:sterileRate}
\end{eqnarray}
where $m_{s}$ is the sterile neutrino mass and $\theta$ is the mixing angle between the active and sterile states.

Limits on $\sin^2 2\theta$ depend on both the observed flux and the fraction of dark matter comprised by sterile neutrinos. For simplicity, limits are typically quoted under the assumption that sterile neutrinos comprise all of the dark matter. The X-ray fluxes reported by the claimed detections discussed above in stacked galaxy clusters, M31, Perseus, and the MW GC correspond roughly to $\sin^2 2\theta$ of $10^{-11}$~to~$10^{-10}$. 

\subsection{Dark Matter Signal for Large FOV Observations}
\label{sec:signal}

\begin{figure}
\begin{center}
\includegraphics[width=\columnwidth]{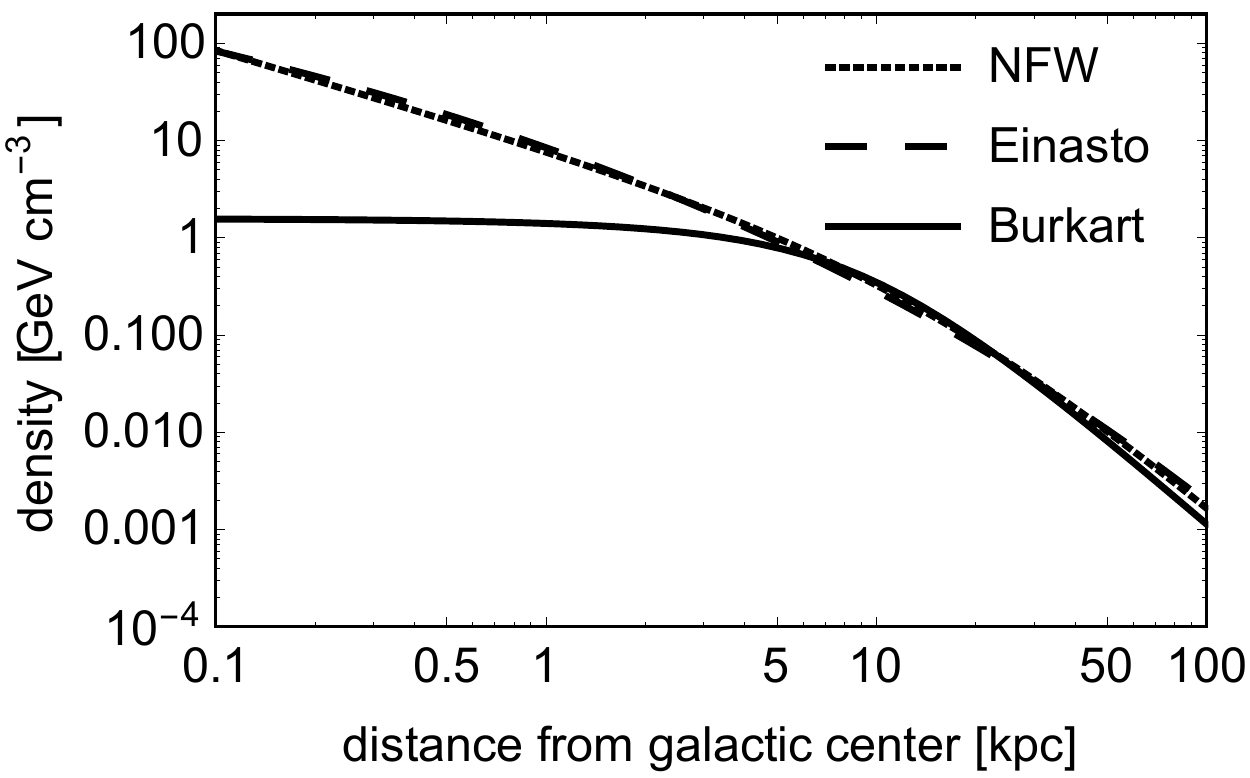}
\end{center}
\caption{Example of common DM halo profiles using NFW \citep{Nesti2013} \emph{(dotted)}, Einasto \citep{Bernal2012} \emph{(dashed)}, and cored Burkart \citep{Nesti2013} \emph{(solid)} parameterizations. \label{fig:DMprofiles}}
\end{figure}


The flux expected from decay of sterile neutrinos in the MW halo is proportional to the integral of the DM density along the line-of-sight and over the field of view
\begin{equation}\label{eqn:flux}
\mathcal{F} = \frac{\Gamma}{m_s} \frac{1}{4\pi} \int_{FOV}  \int_{0}^{\infty} \rho(r(\ell, \psi))~d\ell \, d\Omega,
\end{equation}
where in the integral of the dark matter profile density $\rho(r)$, the parameter $\ell$ is the distance along the line of sight, $r$ is the distance from the GC, and the angular integral is taken over the field of view of the instrument. The distance from the GC is related to the line-of-sight distance by
\begin{equation}
r(\ell, \psi) = \sqrt{\ell^2 + d^2 - 2 \ell d \cos \psi},
\end{equation}
where $d$ is the distance of the earth from the GC and $\psi$ is the opening angle from the GC.


In order to obtain a large FOV, the sounding rocket observations considered in this paper do not use an X-ray optic. The detector observes a field determined by an optical stop and has no imaging capability. The effective area is that of the detector itself, on the order of 1~cm$^{2}$. Furthermore, sounding rocket flights observe for a few hundred seconds per flight. In comparison, \XMM\ has made observations on the order of a megasecond, with an effective area at 3.5~keV of around 200~cm$^{2}$ for each MOS detector. 

\begin{figure}
\begin{center}
\includegraphics[width=\columnwidth]{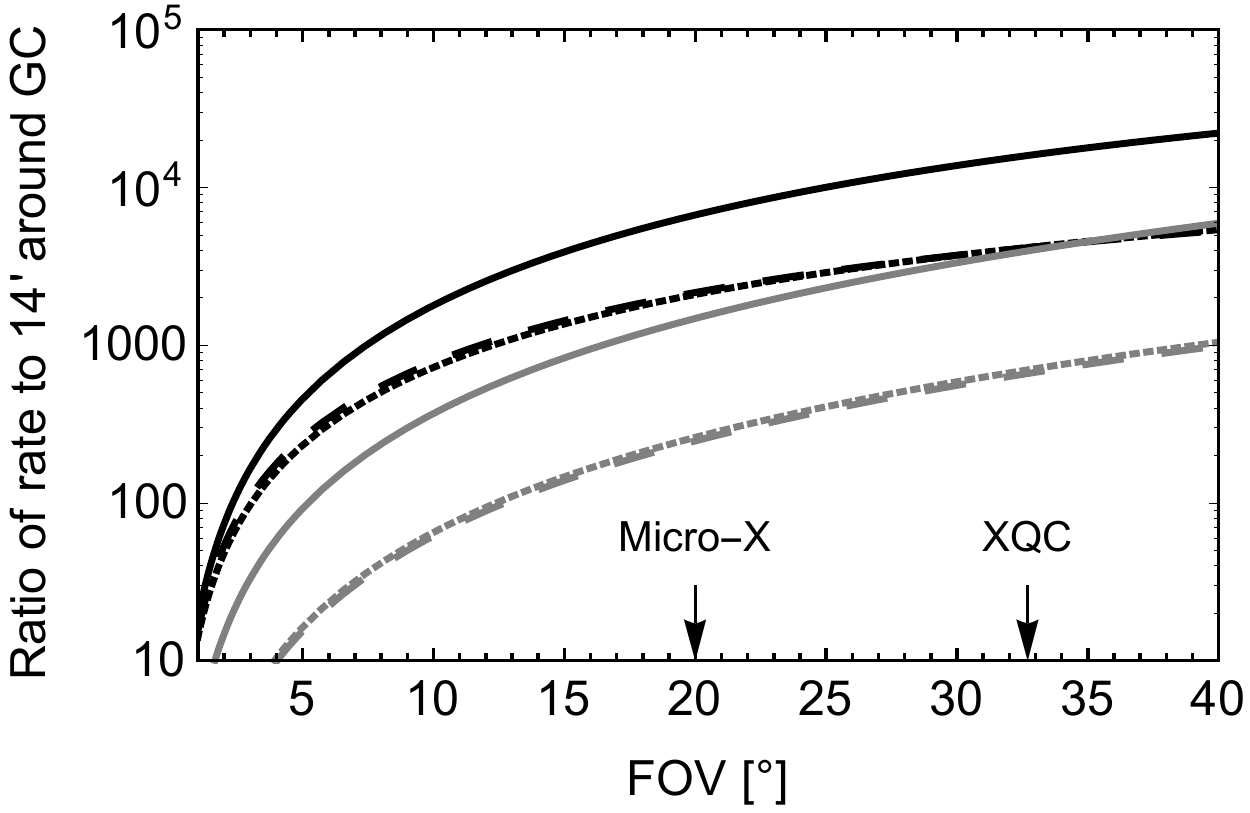}
\end{center}
\caption{Ratio of sterile neutrino decay signal in a large FOV to a 14'-radius FOV (\XMM) around the galactic center as a function of the FOV half-opening angle. Black curves are for an observation centered at the GC, while gray curves are for an observation centered on the XQC field of $(b,l) = (165^\circ, -5^\circ)$. Line style corresponds to different DM profiles: NFW \citep{Nesti2013} \emph{(dotted)}, Einasto \citep{Bernal2012} \emph{(dashed)}, and cored Burkart \citep{Nesti2013} \emph{(solid)} parameterizations. The FOVs for Micro-X and XQC are indicated by arrows on the horizontal axis.\label{fig:FluxRatios}}
\end{figure}

In order to compare between different FOV observations, a DM halo must be assumed, and we show several representative profiles in Figure~\ref{fig:DMprofiles}. In Figure~\ref{fig:FluxRatios} we show in black the ratio of the expected rate from sterile neutrino decay in the central 14' radius of the GC (\XMM 's FOV) to a larger FOV also centered on the GC. In gray we show the ratio between the \XMM\ GC observation and a different field near the MW anti-center at Galactic coordinates $l = 165^\circ, b = -5^\circ$, observed by the 5th flight of the XQC (discussed in the next section). With a sufficiently large FOV, signal rates (in cts~cm$^{-2}\, \mathrm{s}^{-1}$) 3 to 4 orders of magnitude larger than the \XMM\ GC observation are attainable. The background (in this case meaning all X-ray flux of non-DM origin) of observations that include the GC do not increase as quickly with FOV since the X-ray flux from the GC and the Galactic ridge are much stronger than the cosmic X-ray background (CXB) but extend to roughly $\pm 5^\circ$ from the plane. This gives large FOV observations of the GC a better signal-to-noise ratio. Finally, the higher energy resolution of the microcalorimeter instruments in sounding rockets can cut the continuum background per energy bin by over an order of magnitude when compared to the CCD energy resolution of \XMM.

\begin{table}
\begin{center}
\begin{tabular}{l c}
\hline
reference flux (in 14' of GC)		& $2.9 \times 10^{-5}\, \mathrm{cm}^{-2}\, \mathrm{s}^{-1}$ \\
scaled flux (in $20^{\circ}$ of GC)	& $6.1 \times 10^{-2}\, \mathrm{cm}^{-2}\, \mathrm{s}^{-1}$\\
effective area at 3.55~keV			& 1~cm$^{2}$ \\
exposure time					& 300~s \\
resolution (FWHM)				& 3~eV \\
signal events	(in $20^{\circ}$ of GC)	& 18.2 \\
\hline
bg. rate at 3.55~keV (see \S \ref{sec:MicroXBackground})	& $4.5\, \mathrm{cm}^{-2}\, \mathrm{s}^{-1}\, \mathrm{keV}^{-1}$ \\
bg. events in signal window & 6.7 \\
\hline
median signal significance			& $5.6~\sigma$ \\	
\hline
\end{tabular}
\caption{Basic signal and rates expected for a hypothetical observation of the GC using a microcalorimeter on a sounding rocket, assuming a fiducial signal flux from \cite{Boyarsky2014a}.}\label{tab:uxrates}
\end{center}
\end{table}

To get a feel for the potential sensitivity of these observations, consider a hypothetical instrument with a 1~cm$^{2}$ effective area, 3~eV FWHM resolution, and 20$^{\circ}$ radius FOV at 3.55~keV, which observes the GC for 300~s (see Table~\ref{tab:uxrates}). If the flux reported by \cite{Boyarsky2014a} over a 14' radius FOV is due to decaying sterile neutrino dark matter with the NFW profile of Figure~\ref{fig:DMprofiles}, then our hypothetical 20$^{\circ}$ field would expect a scaled signal flux of $6.1 \times 10^{-2}\, \mathrm{cm}^{-2}\, \mathrm{s}^{-1}$, 2000 times higher than the \XMM\, observation. A 300~s observation would measure 18.2 total events in the X-ray line. At 3.5~keV, the background model described in Section~\ref{sec:MicroXBackground} predicts a flux of 4.5~cts~cm$^{-2}\, \mathrm{s}^{-1}\, \mathrm{keV}^{-1}$. The background in a 5.1~eV ($\pm2\sigma_E$) window would be 6.7 events. In spite of the small statistics, the median significance of the putative signal above the continuum background would be $5.6\sigma$ in this short observation.

For comparison, in the \cite{Boyarsky2014a} analysis of 1.4~Ms of \XMM\ data we estimate around 7,500 signal counts in the claimed 3.54~keV line in each MOS detector. In that same resolution element, there are upwards of 500,000 background counts.  With a signal-to-noise ratio of 0.015, the authors use the \XMM\ high statistics measurement to detect such a small signal at high formal significance ($5.7\sigma$), but doing so depends on an accurate model of their background and minimal systematic errors.

%% file: XQClimit.tex
\section{Analysis of XQC Data}
\label{sec:XQCdata}

\begin{figure}
\begin{center}
\includegraphics[width=\columnwidth]{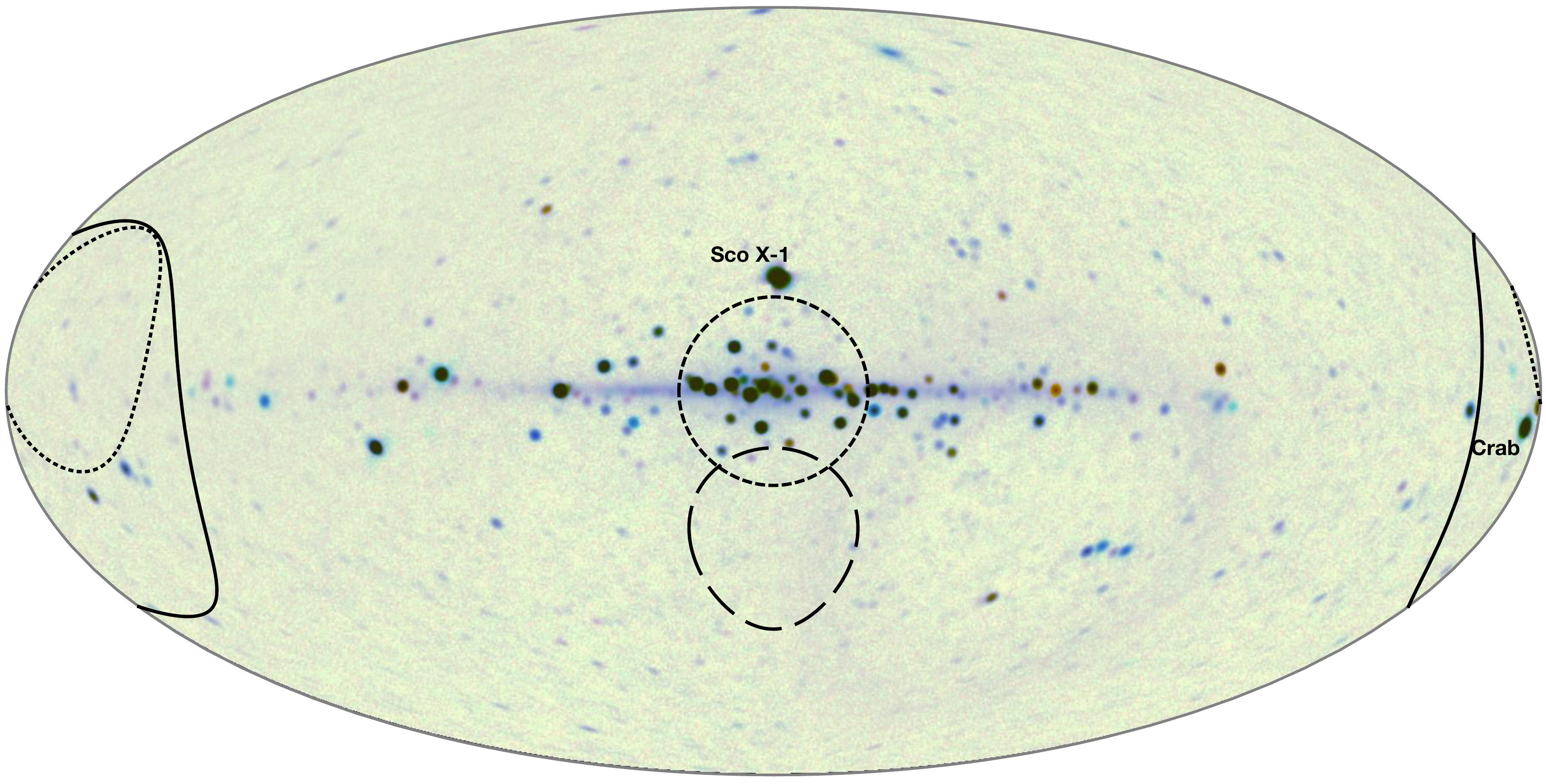}
\end{center}
\caption{All-sky X-ray map from the MAXI/GSC instrument onboard the International Space Station \citep{Mihara2014}. The image is a negative made from a color rendition with the following energy band definitions: red, 2-4~keV, green, 4--10~keV, and blue, 10--20~keV. The XQC field analyzed in this section centered on $l = 165^\circ, b = -5^\circ$ is delineated by the solid line. The dotted line is a Micro-X field centered on $l = 162^\circ, b = 7^\circ$, chosen to lie inside the XQC field and evade the Crab pulsar. The dashed line is the Micro-X GC field and the long-dashed line is the Micro-X off-plane field centered on $l = 0^\circ, b = -32^\circ$, both discussed in Section~\ref{sec:future}. \label{fig:field map}}
\end{figure}

\begin{figure*}
\begin{center}
\includegraphics[width=\textwidth]{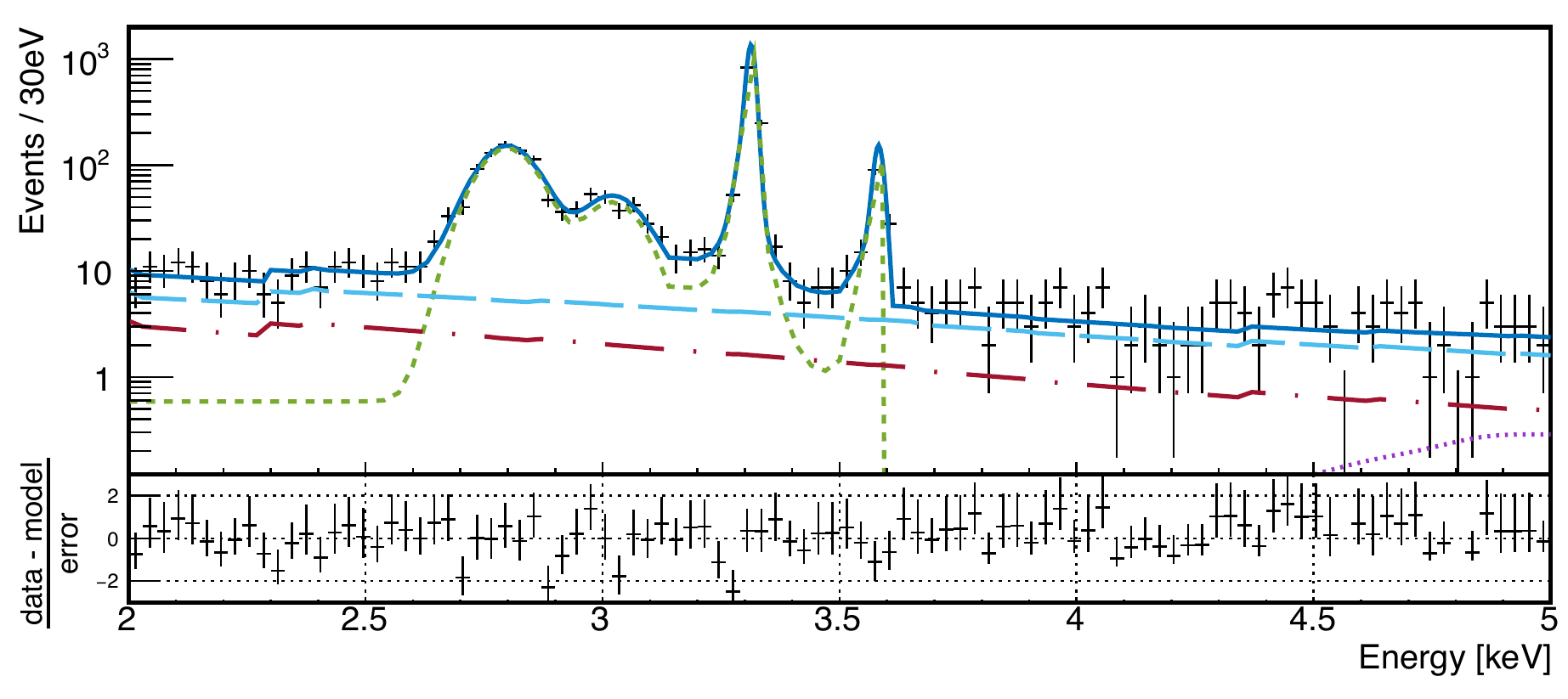}
\end{center}
\caption{Spectrum of XQC data overlaid with fitted total background model \emph{(solid blue)}. Dashed lines show background model components, consisting of a power law continuum from the diffuse X-ray background \emph{(long dashed cyan)}, a power law continuum from the Crab \emph{(dot-dashed red)}, cosmic rays \emph{(dotted purple)}, and lines from the $^{41}$Ca calibration source onboard the instrument \emph{(short dashed green)}. The calibration source produces lines at 3.31~keV and 3.59~keV from K$\alpha$ and K$\beta$ transitions of potassium, while the two broad peaks at lower energies are due to K$\alpha$ and K$\beta$ X-rays that interact in the Si substrate of the HgTe absorbers and experience energy losses due to charge trapping. The flat continuum visible below 2.5~keV in the calibration spectrum is due to source events in which the photoelectron escapes the absorber. The bottom panel shows residuals between the data and the total background model, normalized by the error for each bin. \label{fig:XQCModelFit}}
\end{figure*}

Having laid out the basic strategy, we now focus on existing data from X-Ray Quantum Calorimeter. The XQC payload is a mature flight system with 6 flights between 1995 and 2014 \citep{Crowder2012}. The XQC spectrometer 
is an array of 36 microcalorimeters using ion-implanted semiconductor thermistors each coupled to a 2~mm~$\times$~2~mm$~\times~$0.96~$\mu$m HgTe absorber on a 14~$\mu$m-thick Si substrate, with total area of 1.44~cm$^{2}$. The energy resolution below 1~keV is 11~eV FWHM, although due to position dependence the resolution degrades to 23~eV FWHM at 3.3~keV. The microcalorimeter array is mounted inside a cryogenic system which uses pumped He as a $1.5$~K bath for an Adiabatic Demagnetization Refrigerator (ADR), which is coupled to the detector assembly and cools it to $\sim50$~mK temperatures. To survive the launch vibrations while cold, the cryogenic system is suspended with vibration insulators from the skin of the rocket, and the resonant frequencies of the system are designed to minimize coupling of skin vibrations to the detectors during launch. The FOV of XQC is 1 sr, subtending a 32.3$^{\circ}$ radius in the sky. 

Data from an observation centered at the Galactic coordinates of $l = 90^\circ, b = 60^\circ$ during the 3rd flight of XQC were first presented by \citet{McCammon:2002p160}. \citet{Boyarsky2006a} used this data to constrain the decay of sterile neutrino dark matter, and their results are shown in Figure~\ref{fig:sterileLimits}. Their analysis did not perform background subtraction and was limited to data below $\sim 1$~keV. 

We perform a new analysis that develops a background model for the data between 2.0~keV and 5.0~keV, and then uses the data and background model to constrain the flux of an unidentified line in this interval. The use of background subtraction and higher-energy data from a more recent flight of XQC are the main improvements over \citet{Boyarsky2006a}. We analyze a partial data set from the fifth flight of the XQC rocket, which flew 2011 November 06 at 08:00 UT as flight 36.364UH from the White Sands Missile Range. It obtained about five minutes of on-target data at altitudes above 160 km.  The field of view was centered at the Galactic coordinates of $l = 165^\circ, b = -5^\circ$ (shown in Figure~\ref{fig:field map}), close to the galactic anti-center and including the Crab nebula. A total of 200~s of on-target data was analyzed on 29 functional pixels. After a very conservative quality cut to remove pixels and time periods with unstable event rates, 2551~pixel~$\cdot$~s remain on 24 pixels, for an effective exposure of 106~s per pixel. Data from other XQC flights are also being reprocessed, and the combination of these data sets will increase the total exposure by a factor of a few in a future analysis.

\begin{figure}
\begin{center}
\includegraphics[width=\columnwidth]{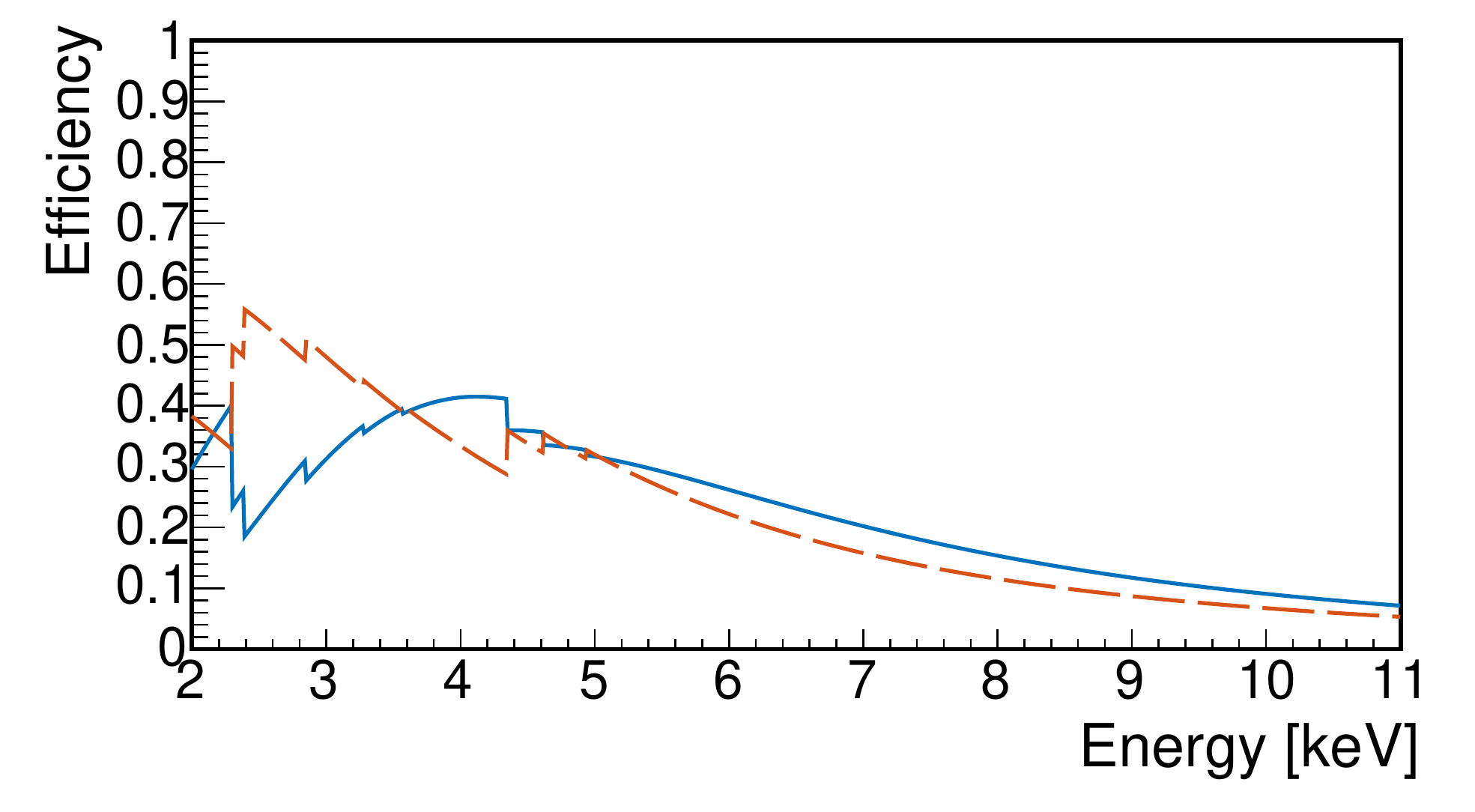}
\end{center}
\caption{Efficiency of XQC for detecting X-rays as a function of energy. Curves show the efficiency of absorption in the HgTe absorber \emph{(dashed orange)}, and the efficiency of absorption in the Si substrate \emph{(solid blue)}. Events absorbed in the Si substrate lose 15-20\% of their energy due to charge trapping, but otherwise appear as good pulses (see text for discussion). \label{fig:XQCEff}}
\end{figure}

Figure~\ref{fig:XQCModelFit} shows the XQC data above 2.0~keV. The spectrum contains a power law continuum with strong lines at 3.31~keV and 3.59~keV from K$\alpha$ and K$\beta$ transitions of potassium, respectively. These lines arise from a $^{41}$Ca source which provides continuous calibration during the flight, and is used to correct gain fluctuations. X-rays incident on XQC may be absorbed in either the HgTe absorber or its Si substrate. The photons absorbed in the HgTe are efficiently thermalized, while those absorbed in the substrate experience an energy loss of about 15-20\%. Potassium K$\alpha$ and K$\beta$ events in the absorber's Si substrate form the two broad peaks centered at 2.8~keV and 3.0~keV, below the corresponding lines due to absorption in the HgTe. The relative intensity of the full-energy peak in the absorber and the second peak from events in the substrate is determined by the relative absorption efficiencies for X-rays in the two detector elements, shown in Figure~\ref{fig:XQCEff}. XQC was optimized for studying the soft X-ray background in the 0.1-1~keV energy range, where the HgTe absorber has higher efficiency and almost all X-rays are absorbed before reaching the Si substrate. A detector with thicker HgTe absorbers would be more suitable for the 2-5~keV region that we study here. The efficiency of the Si rises rapidly above 1.0~keV, and becomes comparable to the HgTe efficiency above 3.5~keV. 

\begin{figure*}
\begin{center}
\includegraphics[width=\textwidth]{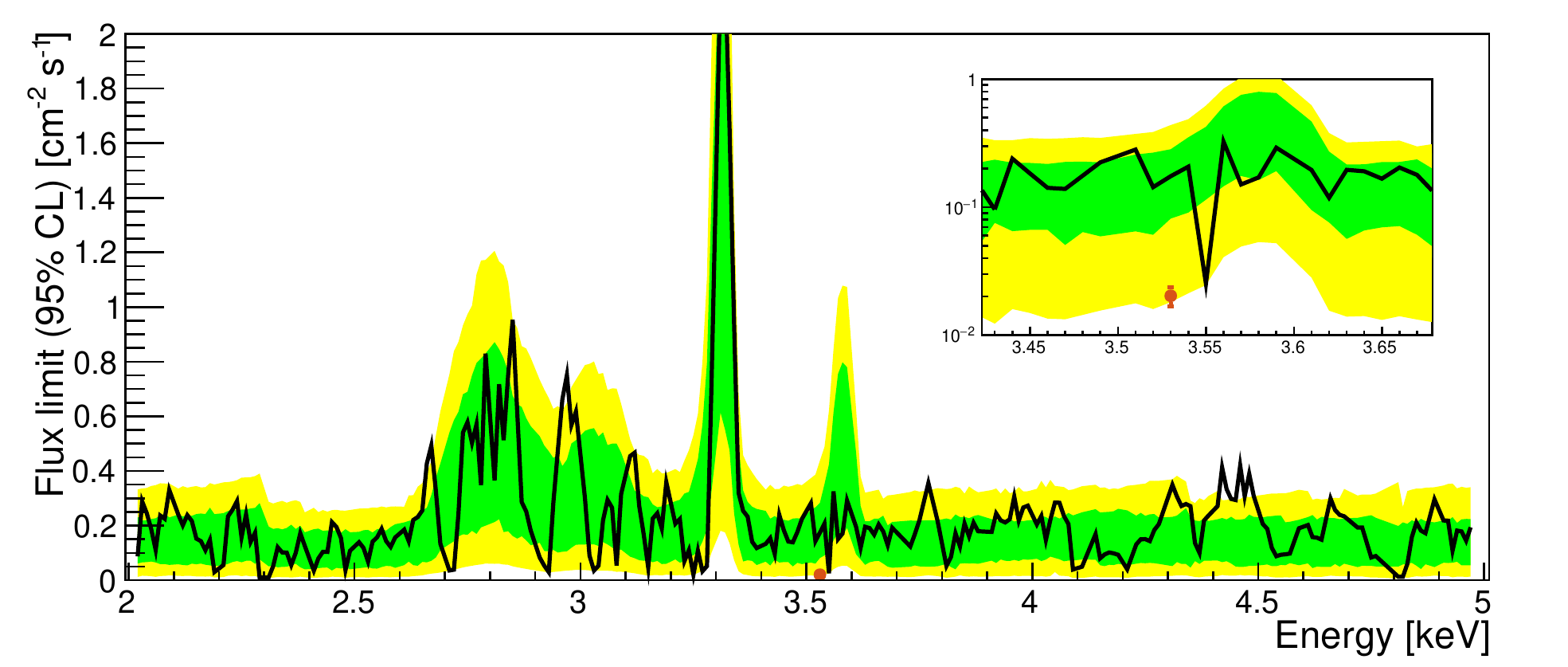}
\end{center}
\caption{Upper limit at 95\% CL on the flux of an unidentified line in the XQC spectrum \emph{(black line)}, as a function of the line energy. Bands show the $\pm 1\sigma$ \emph{(green band)} and $\pm 2\sigma$ \emph{(yellow band)} range of the expected limit from the background-only hypothesis. Inset shows the energy range around 3.5~keV. The flux from the \cite{Boyarsky2014a} claim referred to the XQC field of view using the NFW profile from Figure \ref{fig:FluxRatios} is shown with a red dot. \label{fig:XQCLimits}}
\end{figure*}

The overall strategy of the analysis is to perform an exclusion on the rate above background of an unidentified line centered at each energy between 2.0~keV to 5.0~keV. To do this, we first fit a background model to the data incorporating the important spectral features described above. We then use the model to estimate the background in a sliding energy window, allowing us to set an upper limit on the expected flux from an unidentified line, as a function of energy. Modeling of the energy spectrum becomes more complex at energies below 2.0~keV due to weak atomic lines from thermal emission. At energies significantly above 5~keV the energy scale could become nonlinear and the detection efficiency drops due to saturation. For simplicity, we restrict to a conservative energy window of 2.0-5.0~keV, although this range could likely be expanded in future analyses.

The background model and its components are shown in Figure~\ref{fig:XQCModelFit}. The model incorporates the lines from the $^{41}$Ca source, a continuum from events in which photoelectrons escape the absorber, a power law continuum from the Crab \citep{Mori2004}, a power law continuum from the cosmic X-ray background (CXB) \citep{Hickox2006}, and a component from cosmic rays. The power law models are properly corrected for the suppressed energy measurement when X-rays interact in the Si substrate, as well as for the efficiency of X-ray detection shown in Figure~\ref{fig:XQCEff}. Fits are performed using the RooFit software package based on the Minuit numerical minimizer \citep{Verkerke2003}. We use the method of extended unbinned maximum likelihood \citep{Barlow1990}, which is more stable than binned fits in a low-statistics setting. More details on the model and statistical methodology are provided in the Appendix.

For each energy $E_0$, we construct a window $[E_0 - 2\sigma, E_0 + 2\sigma]$ of four standard deviations in energy resolution which we use to set upper limits on any flux above the modeled background. The background model is fit to all data outside this signal window, and then extrapolated into the window. The background model in the window is then integrated to obtain the background rate $b$ with uncertainty $\sigma_b$ propagated from the uncertainty on the fit parameters. An upper limit is then set on the rate of signal above background in the window, for a Poisson process. Limits are set using the profile likelihood test statistic described in \cite{Cowan2011}, which incorporates the uncertainty on the background rate in the window. The critical value of the test statistic for the desired confidence level is calculated exactly from Monte Carlo simulations rather than using the asymptotic distribution of the test statistic. This is important at higher energies, where the low statistics cause the test statistic to differ from its asymptotic distribution. The upper limits are set using the RooStats software package \citep{Moneta2011}. Although the ``sliding window" approach does not exploit the signal and background shape within the energy window, there is little loss of information because of the very low statistics of a potential signal relative to the slowly-varying background within the narrow window.

Figure \ref{fig:XQCLimits} shows the limits on the flux of an unidentified line as a function of the line energy, calculated using the limit-setting procedure described above with the background model. At 3.53~keV, we set an upper limit on the flux of an unidentified line of $0.17$~cm$^{-2}$~s$^{-1}$ at 95\% CL. The flux reported in the GC by \cite{Boyarsky2014a}, when referred to the XQC field using fiducial DM profiles is listed in Table \ref{tab:XQClimits}. The inset of Figure \ref{fig:XQCLimits} also shows a strong downward fluctuation of the limit at 3.55~keV, consistent with the lower edge of the $\pm2\sigma$ band of expected limits. While this fluctuation may be purely random, it could also be caused by a very slight difference in the lower tail of the K$\beta$ line between the flight data and the calibration template constructed using ground calibration data. Future flights of XQC would clearly benefit by choosing a calibration source with lines further from the signal region near 3.5~keV. Although the XQC data do not exclude the Boyarsky best-fit flux, the upper limit is close in the case of a cored DM profile, such as Burkart. It is furthermore important to emphasize that XQC achieves this result with merely $\sim106$~s of data on only a subset of its pixels. This underscores the value of combining additional datasets from other flights: additional statistics will both improve the limit and enable more robust background modeling.

\begin{table}
\begin{center}
\begin{tabular}{| l | l | l |}
\hline
Data									& DM profile 			& Flux [cm$^{-2}$~s$^{-1}$]\\
\hhline{|=|=|=|}
XQC (this work)					& N/A						& $<17 \times 10^{-2}$ (95\% CL)  \\
\hline
Expected line flux in 				& NFW					& $(2.03 \pm 0.35) \times 10^{-2}$  \\
XQC field scaled from 					& Einasto				& $(1.95 \pm 0.34) \times 10^{-2}$  \\
\cite{Boyarsky2014a}	& Burkart				& $(11.7 \pm 0.2) \times 10^{-2}$  \\
\hline
\end{tabular}
\caption{Flux limits on a line at 3.53~keV from XQC data in this work, compared with the expected flux in the XQC field obtained by scaling the galactic center observation of \cite{Boyarsky2014a}, using the DM profiles of Figure~\ref{fig:FluxRatios}. \label{tab:XQClimits}}
\end{center}
\end{table}

%% file: backgrounds.tex
\section{Estimates for Future Observations}
\label{sec:future}

The XQC limits shown in the previous section are photon-limited; more observation time will result in better sensitivity. We also plan future observations in several fields including the GC to both increase sensitivity to potential undiscovered sterile neutrino lines and provide a definitive test of the nature of the line found by \cite{Boyarsky2014a}. These future observations will benefit from higher resolution microcalorimeter arrays, such as those developed for the Micro-X rocket payload. 

\begin{figure}
\begin{center}
\includegraphics[width=\columnwidth]{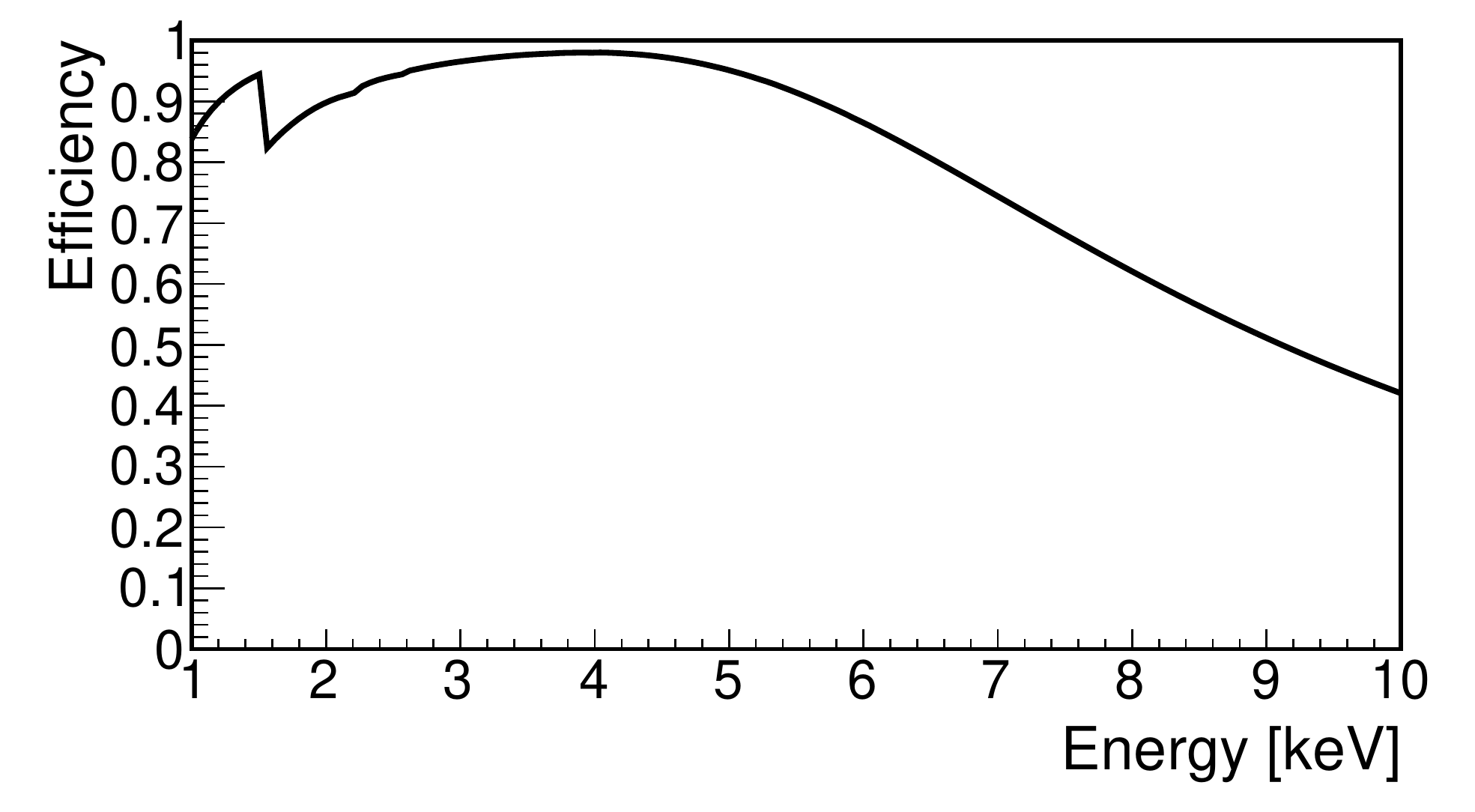}
\end{center}
\caption{Total efficiency of Micro-X for X-ray detection as a function of energy. The total area of the detector is 1~cm$^{2}$.\label{fig:uXeff}}
\end{figure}

The Micro-X payload is a new system based on the XQC design, with new detectors and readout to allow for a larger array of higher-resolution microcalorimeters \citep{Heine2014}. Although Micro-X was designed to be used with a 2.1~m X-ray optic, for dark matter searches it will be reconfigured to fly without it, using an optical stop like XQC to set its field of view. 
The Micro-X detector consists of an array of 128 microcalorimeters using Transition-Edge Sensor (TES) thermometers each coupled to a 0.6~mm~$\times$~0.6~mm BiAu absorber. For the dark matter flight, new absorbers with 0.9~mm per side and thickness of 3~$\mu$m (Bi) + 0.7~$\mu$m (Au) will be used, with total area of 1~cm$^{2}$. The absorption efficiency of X-rays in the the detector is shown in Figure~\ref{fig:uXeff}. The BiAu absorbers allow for better thermalization and should minimize position dependence, allowing Micro-X to retain its design resolution of 3~eV FWHM at 3.5~keV. 

In this section we estimate the sensitivity of a potential GC observation with the Micro-X payload. A GC observation is the most direct comparison to \cite{Boyarsky2014a}, and the higher energy resolution of the Micro-X instrument results in a lower background from continuum emission and better discrimination between unexpected lines and those coming from atomic transitions in the observed plasma. Micro-X is designed to be coupled to a mirror, so for this calculation we use a 0.38~sr, $20^{\circ}$ radius FOV as an estimate of the achievable FOV without the Micro-X optics. The Micro-X GC field is shown in Figure~\ref{fig:field map}. A future redesign of the optical aperture of the cryostat could increase the FOV to 1~sr. 

\subsection{Galactic Center Backgrounds}
\label{sec:MicroXBackground}

\begin{figure}
\begin{center}
\includegraphics[width=\columnwidth]{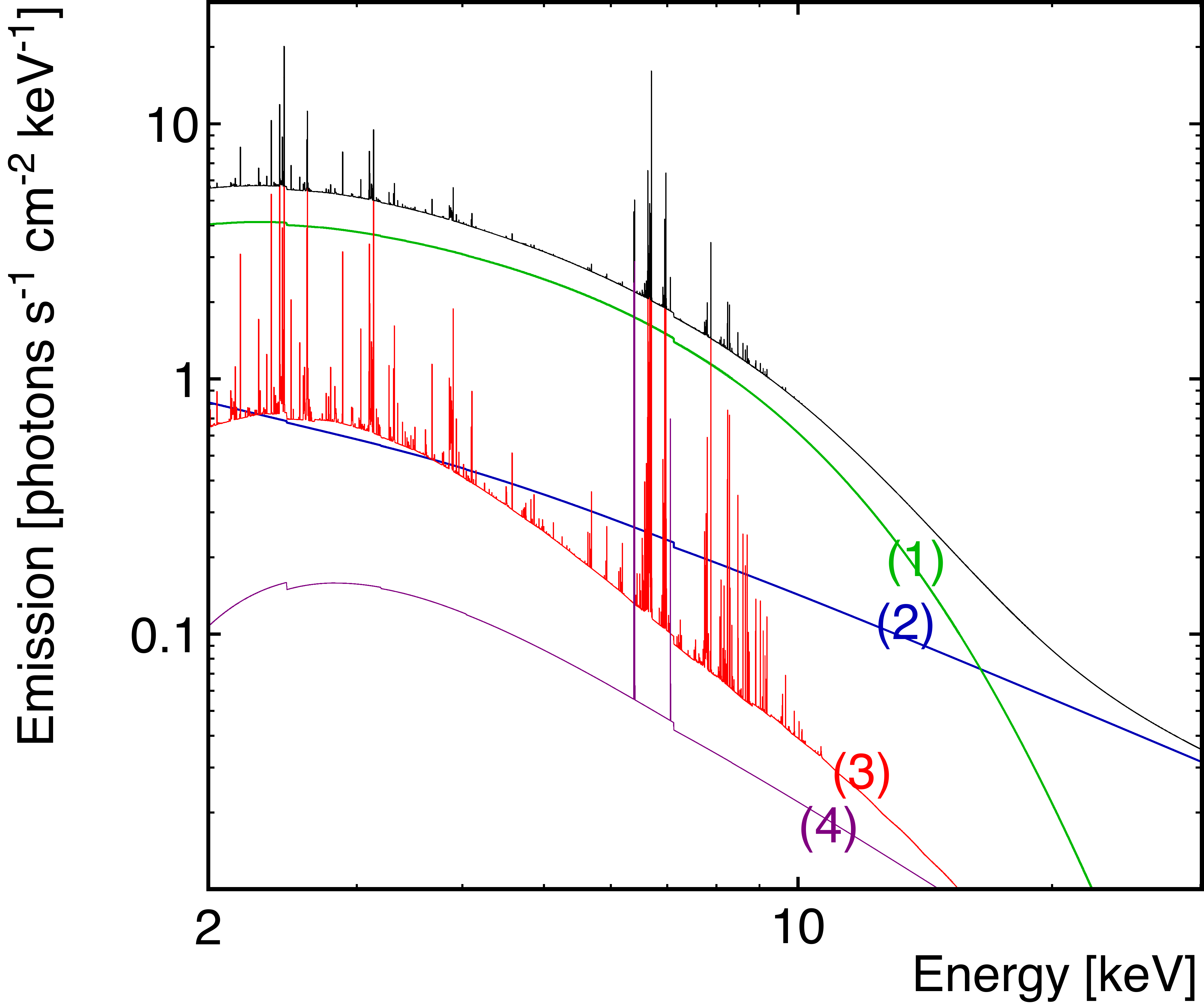}
\end{center}
\caption{Expected X-ray background from a 20$^\circ$ radius region centered on the Galactic Center. The total background spectrum is shown in black. The expected emission from the brightest low mass X-ray binaries (as shown in Table \ref{tab:uxbkg}) is shown in green and labeled (1),  the emission from the cosmic X-ray background is shown in blue and labeled (2), the thermal components from the galactic diffuse background emission are shown in green and labeled (3), and finally the purple line labeled (4) shows the contribution from ionized cold ISM neutral Fe, which is a combination of a powerlaw continuum and gaussians representing the K$\alpha$ and K$\beta$ iron line emission. The energy bins are 3~eV wide.}
\label{fig:uxbkg}
\end{figure}

\begin{figure}
\begin{center}
\includegraphics[width=\columnwidth]{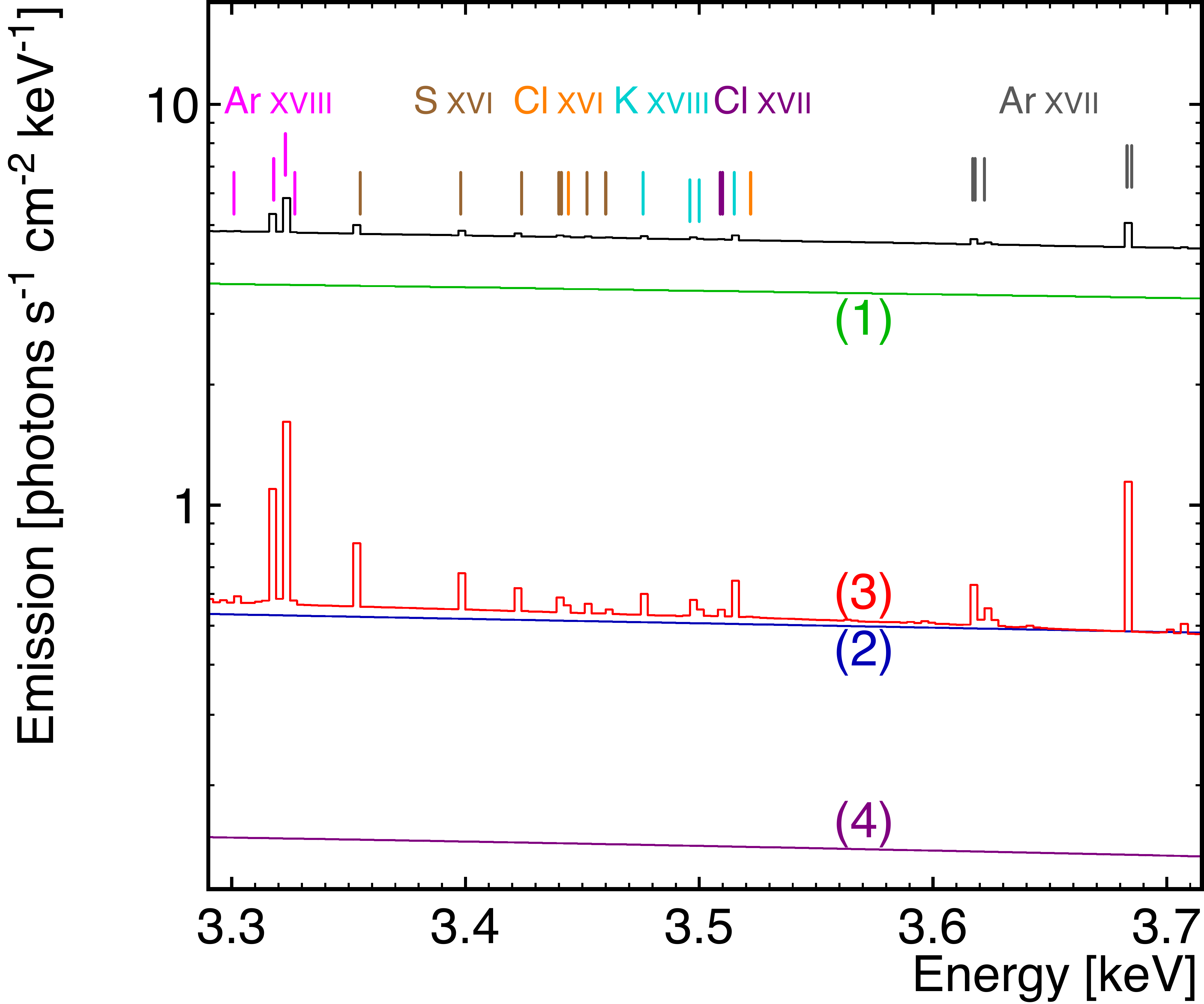}
\end{center}
\caption{Background from a 20$^\circ$ radius region centered on the GC (as in Figure~\ref{fig:uxbkg}), in the 3.3-3.7 keV range. The total background spectrum is shown in black, and the components are as in Figure~\ref{fig:uxbkg}. Emission lines from ions Ar {\scshape xviii}, S {\scshape xvi}, Cl {\scshape xvi}, Cl {\scshape xvii}, K {\scshape xviii}, and Ar {\scshape xvii} are also shown. The abundance of Cl and K have been set to solar values, using the tables from \citet{anders_1989}. The energy bins are 3~eV wide.}
\label{fig:uxbkg-lines}
\end{figure}

The first step in estimating the sensitivity of a potential MW GC observation with Micro-X is constructing a background model of the complex emission from the large FOV. For this we have used results from {\it Suzaku} observations of the Galactic Ridge (GR) and GC to estimate the contribution from thermal diffuse emission \citep{uchiyama_2013}, the cosmic X-ray background (CXB) component from unresolved extragalactic sources as modeled by \citep{kushino_2002} from {\it ASCA} observations, and the \ROSAT\ All Sky Survey -- Bright Source Catalogue \citep[RASS-BSC, revision 1RXS,][]{voges_1999} to estimate the contribution from point sources in the field. The background model shown in Figures~\ref{fig:uxbkg} and \ref{fig:uxbkg-lines} is discussed below.

\subsubsection{Diffuse Background}

The spectral model for the diffuse emission expected from the field is constructed using the thermal components from observations of the GC and GR taken with Suzaku \citep{uchiyama_2013}, as well as the isotropic CXB model from \citet{kushino_2002}. The GC component is defined as emission from a region with radius $0.6^{\circ}$, centered at $l,b=0^{\circ},0^{\circ}$. The GR region extends from $-5^{\circ}$ to $+5^{\circ}$ in Galactic latitude and spans the full longitude range of the FOV (from $340^{\circ}$ to $20^{\circ}$). We have defined the spatial extent of these regions using two-exponential model fit to the 2.3-8 keV intensity profile around the GC presented in Table 2 of \citet{uchiyama_2013}. The integrated background spectrum is the sum of the emission from the GC, the GR, and the emission from latitudes $|b|\geq4.9^{\circ}$ to the edge of our FOV. 

Each of the Galactic components is a combination of emission from two thermal plasmas with different temperatures and metallicities, all absorbed through the appropriate column densities. The thermal plasmas are simulated in XSPEC (version 12.8.2l) using the \emph{APEC} model from AtomDB version 3.0.1 \citep{smith_2001,foster_2012}. 
We have limited our analysis to energies above 2.3 keV to avoid the contribution from additional ($|b|>5^{\circ}$) Galactic thermal components from regions extended beyond the GR region, since detailed spatial and spectral models of such extended diffuse thermal emission are currently not available. 
Neutral atoms from the cold ISM are ionized by either low energy cosmic rays or X-ray emission from external sources. This results in an additional diffuse X-ray component from the GC and GR, with a continuum modeled as a powerlaw ($\Gamma=2.13$), and fluorescent K-shell line emission from neutral Fe (shown in purple in Figure~\ref{fig:uxbkg} and Figure~\ref{fig:uxbkg-lines}). We used the normalization of the continuum and equivalent widths of the Fe K$\alpha$ and K$\beta$ lines presented in \citet{nobukawa_2010} and \citet{uchiyama_2013}. However, we split the Fe K$\alpha$ contribution into Fe K$\alpha1$ (at 6.403 keV) and Fe K$\alpha2$ (at 6.390 keV), with 2:1 relative intensities \citep{bearden_1967,kaastra_1993}. 
The CXB contribution is added as a powerlaw spectral component, given by $8.2\times10^{-7} (E/\text{1 keV})^{-1.4}\text{ photons cm}^{-2} \text{ s}^{-1} \text{arcmin}^{-2} \text{keV}^{-1}$, absorbed through the appropriate column density in each direction \citep{kushino_2002}. All spectral parameters are taken from Table 3 of \citet{uchiyama_2013}, and the normalizations of all components are adjusted to match the nominal intensities predicted in that work for the spatial extents described above. The total thermal emission is shown in red in Figure \ref{fig:uxbkg}, and the integrated CXB emission in presented in blue.


\subsubsection{Background from Bright Sources}

\begin{table*}\scriptsize
\begin{center}
\caption{LMXBs included in the background model for the Micro-X observation.}
\label{tab:uxbkg}
\begin{tabular}{ccccccc}
\hline
ROSAT Name & Associated Name & ROSAT Count Rate & $N_{\text{H}}$ & $F_{0.5-2 \text{ keV}}$ & $F_{2-10 \text{ keV}}$ & Reference \\
&&(counts/s) &($10^{22}$ cm$^2$) & \multicolumn{2}{c}{($10^{-9}$ ergs/s/cm$^2$)}&\\
\hline
1RXS J173143.6--165736 & GX 9+9 & 144.5 & 0.15 & 1.3448 & 4.5127 & \citet{ng_2010} \\
1RXS J182340.5--302137 & 4U 1820--30 & 127 & 0.078 & 1.3219 & 5.188 & \citet{costantini_2012} \\
1RXS J170544.6--362527 & GX 349+02 & 91.82 & 0.7 & 1.8785 & 12.075 & \citet{ng_2010} \\
1RXS J173858.1--442659 & 4U 1735--44  & 61.88 & 0.185 & 1.1257 & 4.5544 & \citet{ng_2010} \\
1RXS J175840.1--334828 & 4U 1755--33 & 31.08 & -- & -- & -- & \citet{angelini_2003} \\
1RXS J180132.3--203132 & GX 9+1 & 30.92 & 1.447 & 0.71644 & 19.997 & \citet{iaria_2005} \\
1RXS J173602.0--272541 & GS 1732--273 & 29.79 & 0.67 & 0.82566 & 1.2021 & \citet{yamauchi_2004} \\
1RXS J181601.2--140213 & GX 17+2 & 24.82 & 3.18 & 0.6595 & 15.843 & \citet{cackett_2009} \\
1RXS J170855.6--440653 & 4U 1705--44  & 20.37 & 1.5 & 0.5785 & 6.4055 & \citet{ng_2010} \\
1RXS J174755.8--263352 & GX 3+1 & 19.2 & 1.7 & 0.42601 & 4.4799 & \citet{piraino_2012} \\
1RXS J180108.7--250444 & GX 5--1 & 17.78 & 2.8 & 0.6671 & 41.512 & \citet{ueda_2005} \\
1RXS J173413.0--260527 & KS 1731--260 & 14.09 & 1.08 & 0.40815 & 2.9953 & \citet{narita_2001} \\ 
\hline
\end{tabular}
\end{center}
\end{table*}

The RASS-BSC includes 558 sources within $20^\circ$ of the GC, with a total count rate in the \ROSAT\ broad band (0.1-2.4 keV) of approximately 760 counts$/$s. The 12 brightest sources within the region of interest, all low mass X-ray binaries (LMXBs), account for 80\%  ($\sim613$ counts$/$s) of the total count rate from the resolved bright X-ray source population in the field, and we have listed them in Table \ref{tab:uxbkg}. Their spectra are modeled using the emission and absorption parameters obtained through X-ray observations, using the references also included in Table \ref{tab:uxbkg}. 
Note that the observed flux and spectral shape of these sources will depend on their state at the time of observation, so there is some inherent uncertainty in estimating this background component. 
Since the remaining portion of the X-ray bright source population is also dominated by LMXBs, the integrated spectrum from these 12 brightest sources has been normalized in order for the total flux in the 0.1-2.4 keV band to match the combined \ROSAT\ count rate from all resolved bright sources in the region of interest. 
The integrated BSC spectrum is shown in green in Figure~\ref{fig:uxbkg}. 


\subsubsection{Combined Sky Background}
Figure~\ref{fig:uxbkg} shows the total background model for a $20^{\circ}$ radius field centered on the GC, which includes the diffuse and point source emission discussed above. We show the total flux in black, the expected emission from bright sources (as shown in Table \ref{tab:uxbkg}) in green,  the CXB power-law in blue and thermal components from the Galactic Ridge and Galactic Center in red. 
Figure~\ref{fig:uxbkg-lines} shows the these different components but zooms in on the 3.3 keV to 3.7 keV region (line colors are consistent with those of Figure~\ref{fig:uxbkg}). The continuum in this energy band is dominated by the emission from the BSC LMXBs, yet some thermal line emission is also significant. The relevant ions in this energy range are Ar {\scshape xviii}, S {\scshape xvi}, Cl {\scshape xvi}, Cl {\scshape xvii}, K {\scshape xviii}, and Ar {\scshape xvii}. We used the abundances table of \citet{anders_1989} and set the abundance of Cl and K to those values.

\subsubsection{Instrumental Backgrounds}

In a realistic flight, Micro-X requires an on-board calibration source similar to the one that produces the potassium lines in Figure~\ref{fig:XQCModelFit}. The $^{41}$Ca source used by XQC is not ideal for searching for a line at 3.5~keV because the K$\beta$ line is at an energy similar to the signal. We consider an alternate calibration source consisting of an $^{55}$Fe source that produces fluorescence X-rays by illuminating an NaCl wafer. A kapton filter could block Auger electrons from the NaCl and the $\sim1$~keV X-rays from Na fluorescence, leaving only the K$\alpha$ (2.62~keV) and K$\beta$ (2.82~keV) lines from Cl fluorescence, as well as a small number of back-scattered $^{55}$Fe X-rays. We simulate the energy spectrum observed by Micro-X using Geant4 \citep{Agostinelli2003}, and add it to the astrophysical X-ray spectrum, assuming a calibration source rate of 1~Hz / pixel.

Cosmic rays will also produce a background in Micro-X. Most cosmic ray primaries are protons with energies around a few GeV. These act as minimum-ionizing particles producing a broad continuum of energies in the detector, but peaking in the signal region around 3-4~keV. We simulate the cosmic ray energy spectrum using Geant4, and add this as a background component in the background spectrum. The total rate of cosmic rays is about 1~Hz in Micro-X, which is significantly less than the total rate expected from astrophysical sources.

%% file: Micro-X-sensitivity.tex

\subsection{Signal and Sensitivity Estimates}
\label{sec:MicroXLimits}

\begin{figure}
\begin{center}
\includegraphics[width=\columnwidth]{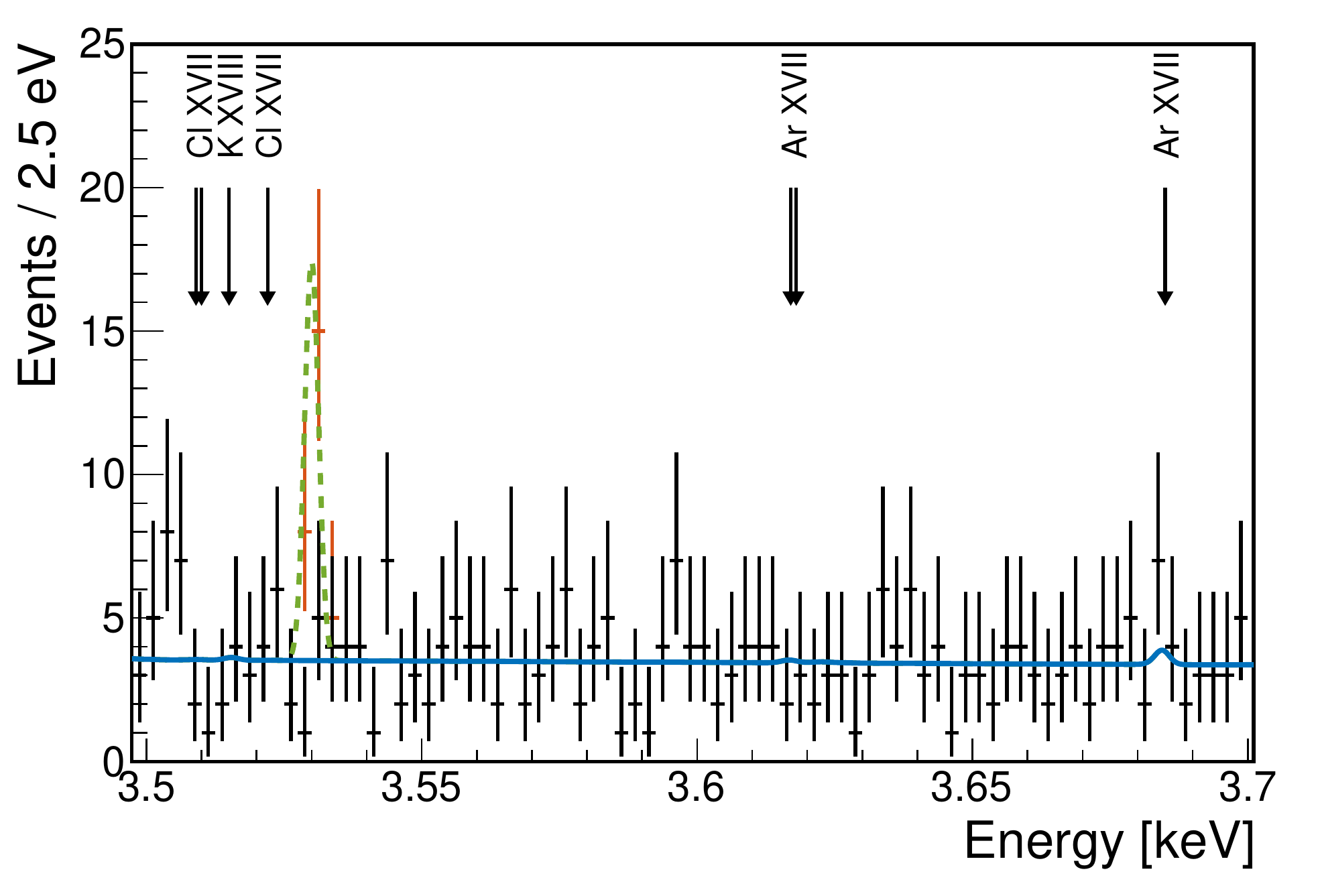}
\end{center}
\caption{Mock data in the energy range of interest for the 3.5~keV line, with \emph{(red)} and without \emph{(black)} a signal, with background model \emph{(blue line)} and signal model \emph{(dashed green line)} overlaid. Note that the excellent spectral resolution of Micro-X provides significant separation of a signal line from nearby atomic lines.\label{fig:uXSignalInjection}}
\label{fig:uxdata}
\end{figure}

A mock observation of the GC with and without the \cite{Boyarsky2014a} line is shown in Figure~\ref{fig:uxdata}. A candidate line of this strength in the Micro-X observation would also be stronger relative to atomic lines than in \XMM\ and \Chandra\ observations. In fact, since the background model for the GC is dominated by a blackbody continuum from low-mass X-ray binaries, no significant atomic lines are expected to be visible above the continuum in the energy range of interest. The strongest lines between 3.5 and 3.6~keV come from K~XVIII and Cl~XVII with expected counts of less than 1 event in the Micro-X observation, so if a line of this strength was detected at this energy it would be in strong tension with a ``standard'' astrophysical origin. As can be seen in Fig~\ref{fig:uxdata}, the putative sterile neutrino line would be bigger than the Ar~XVII line at 3.683--3.685~keV (which is the sum of two emission lines), and this Ar line, in turn, is expected to be a factor of 30 (5) larger than the brightest Cl~XVII (K~XVIII) line.

Using the background model and the methodology of our analysis of XQC data, we can estimate the sensitivity to an unidentified line over the entire energy range available in observations by Micro-X. We consider a 300~s measurement centered on the GC. While the Micro-X constraints are qualitatively similar to those of XQC, they are more stringent because of the significantly higher spectral resolution and exposure of Micro-X.


\begin{figure}
  \begin{center}
  \subfigure{\includegraphics[width=\columnwidth]{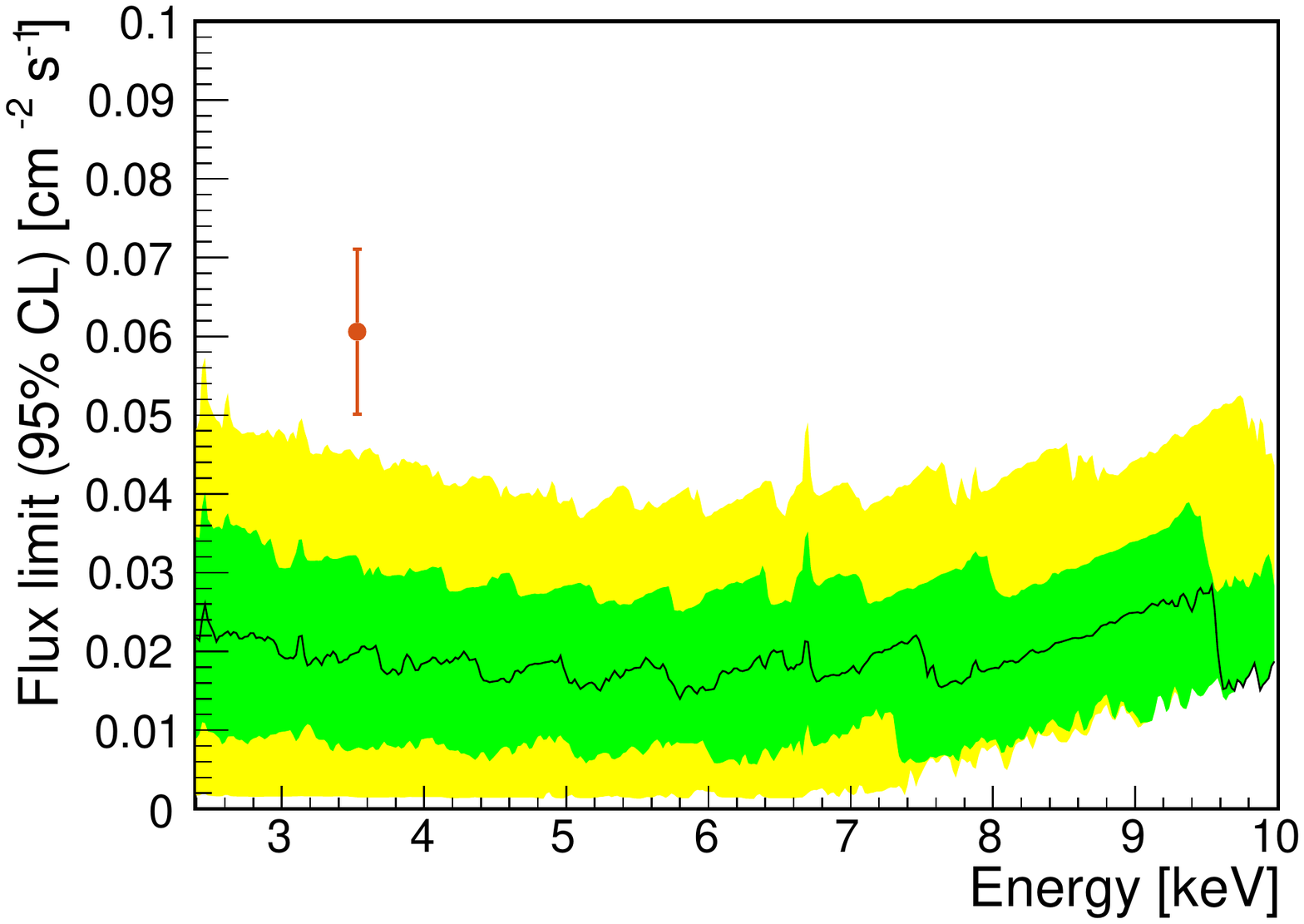}}
  \end{center}
  \caption{Expected limit from an observation of a 20$^\circ$ field around the GC by Micro-X. Black line shows the median expected 95\% CL upper limit, while the green and yellow bands are the $\pm1\sigma$ and $\pm2\sigma$ ranges of the expected upper limits. The expected upper limit rises at high energies because of the falling efficiency to detect x-rays. The red point is the flux of \cite{Boyarsky2014a}, extrapolated to the Micro-X field of view using an NFW profile. Gray bands overlay strong calibration source lines.\label{fig:uXFluxLimit}}
\end{figure}

Using the same analysis approach that we applied to XQC, we compute the expected upper limit on the flux of an unidentified line as a function of energy, under the background-only hypothesis. The resulting limit is shown in Figure~\ref{fig:uXFluxLimit}. Fluctuations of the limit at low energies are due to the presence of atomic lines. At higher energies, large jumps in the limit are caused by the small number of background events in the signal region, while small fluctuations are due to finite statistics in the Monte Carlo simulation used to set the upper limit.


\section{Sterile Neutrino Interpretation}\label{sec:sterilelimit}

\begin{figure*}
\begin{center}
\vspace{24pt}
\includegraphics[width=\textwidth]{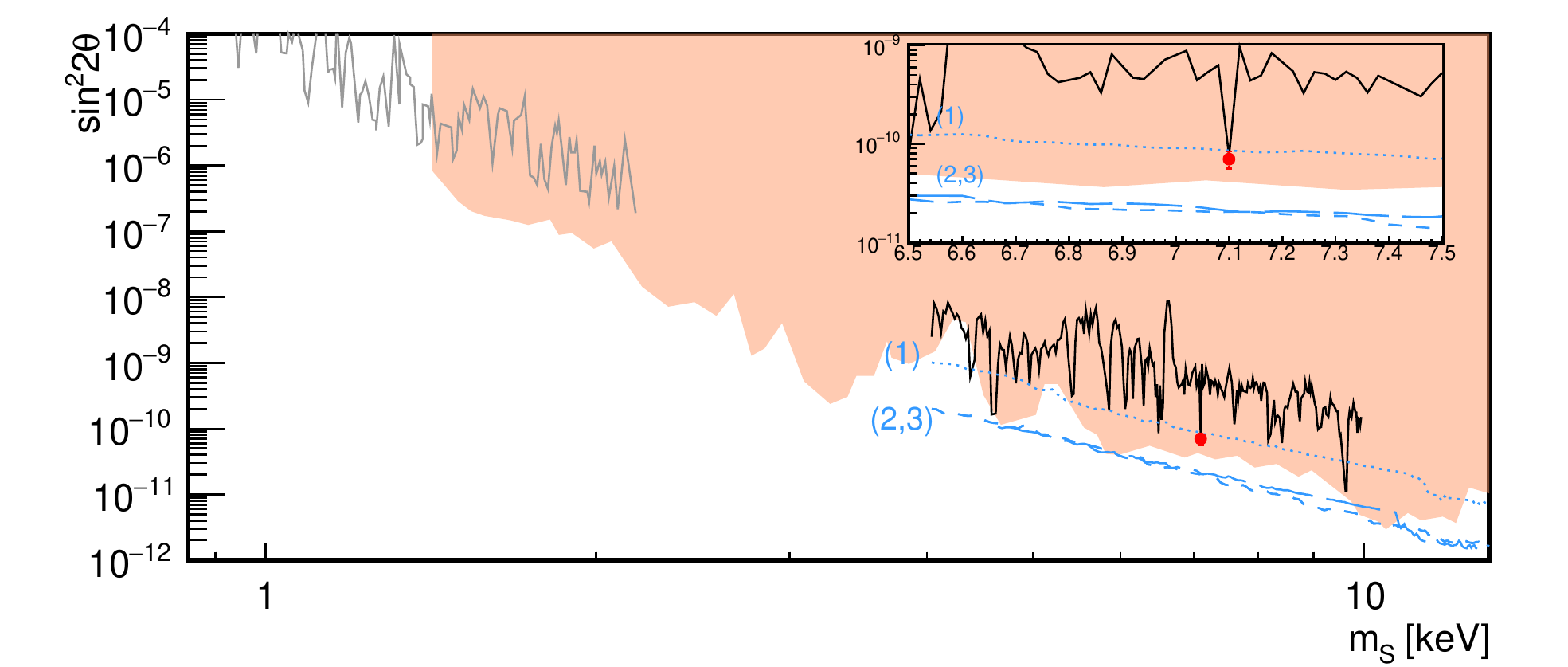}
\end{center}
\caption{Constraints on decaying sterile neutrino dark matter, assuming that sterile neutrinos comprise all of the DM in the NFW profile of \cite{Nesti2013}. Limits include the XQC observation analyzed in this work \emph{(black)}; Micro-X median expectation from an observation of the GC \emph{(1, dotted blue)}, an observation below the plane of the galaxy in the direction of $l = 0^\circ, b = -32^\circ$ \emph{(2, short dashed blue)}, and an observation within the XQC field in the direction of $l = 162^\circ, b = 7^\circ$ \emph{(3, long dashed blue)}; constraints from M31 \citep{Horiuchi2014} \emph{(shaded orange)}; and constraints from the previous analysis of XQC data by \cite{Boyarsky2006a} \emph{(gray)}. The putative signal of \cite{Bulbul2014} is also shown \emph{(red point)}. \label{fig:sterileLimits}}
\end{figure*}

The flux limits obtained from XQC data and projected for Micro-X can be translated into constraints on models of dark matter. Although the literature contains a range of models that could produce an X-ray line, we consider a decaying sterile neutrino as a benchmark model because it has been extensively discussed as the source of the 3.5~keV excess. Using equations (\ref{eqn:sterileRate}) and (\ref{eqn:flux}) for the flux of a decaying sterile neutrino, and assuming an NFW profile, we translate the XQC flux limits of Figure~\ref{fig:XQCLimits} into limits on the sterile neutrino mass $m_s$ and mixing angle $\sin^2 2\theta$, shown as the black line is Figure~\ref{fig:sterileLimits}. 

For the Micro-X payload, we show three sensitivity projections, corresponding to the limits obtained for simulated background-only 300~s observations of the three fields shown in Figure~\ref{fig:field map}. The dotted line labeled (1) in Figure~\ref{fig:sterileLimits} 
shows the limit expected from a Micro-X observation inside of the XQC field at $l = 162^\circ, b = 7^\circ$, chosen to avoid the Crab pulsar. For this projection, we take the background to be the diffuse cosmic X-ray background measured by the XQC observation in Figure~\ref{fig:XQCModelFit}, scaled to the smaller Micro-X field of view, and without the Crab component. The dashed line labeled (2) in Figure~\ref{fig:sterileLimits} shows the limit expected from an observation of the galactic center using the flux limit shown in Figure~\ref{fig:uXFluxLimit}. Finally, the long-dashed line labeled (3) in Figure~\ref{fig:sterileLimits} shows the limit of an off-Galactic-plane pointing at $l = 0^\circ, b = -32^\circ$. For this observation, the background is assumed to be only the cosmic X-ray background shown in blue in Figs.~\ref{fig:uxbkg} and \ref{fig:uxbkg-lines}, reducing the continuum by a factor of $\sim$9 (at 3.5 keV). The putative sterile neutrino signal is reduced by a factor of $\sim$2 relative to the GC assuming an NFW profile, giving this field an improvement in signal-to-noise of 4.5 over the GC pointing. The limit, however, does not improve appreciably over the GC limit. This is due to the fact that the short 300~s observations are photon starved, and the limits are driven by low-number statistics. Thus integrating over multiple flights will yield sensitivity improvements that will increase faster than sqrt(exposure). The off-Galactic-plane fields are particularly appealing since they have a much simpler background with fewer systematics than the GC. In the case of a positive signal, multiple pointings could be used to map out the expected DM profile.

Although the XQC limit is not strong enough to provide a robust exclusion of the parameters inferred by \cite{Bulbul2014}, a Micro-X observation of the GC or a region near the GC could provide a significantly stronger constraint because of its better energy resolution and larger exposure, and because of the larger signal strength in the GC region. These limits obviously depend on the structure of the dark matter halo, with cored profiles producing stronger constraints than NFW-like profiles. 

These wide-FOV rocket observations are complementary to narrow-FOV observations of dwarfs, galaxies, and clusters because they directly address whether an unidentified line is present as an all-sky signal in the MW. This confirmation would be crucial for distinguishing an atomic interpretation from an exotic DM one, and for establishing the signal scaling as a function of the integrated DM density.

%% file: conclusion.tex
\section{Conclusion}
\label{sec:conclusion}

Microcalorimeters onboard sounding rockets have the ability to place competitive bounds on keV sterile neutrinos or other dark matter models whose flux scales linearly with dark matter density. We have analyzed a subset of the data acquired during the 5th flight of the XQC payload corresponding to an effective exposure of 106~s on 24 pixels and placed a upper limit on keV sterile neutrinos between 4-10~keV which demonstrates the prospect for future observations with this type of instrument. A study of future observations in and around the Milky Way galactic center with the Micro-X payload shows that it will have sensitivity to new parameter space in the ($m_{s}$,\,\,$\sin^{2}2\theta$) sterile neutrino plane. Optimizations of the pointing direction, increased field of view and energy resolution, and repeated observations will all increase the sensitivity of this technique in the future.

\acknowledgements
EFF acknowledges support from NASA Award NNX13AD02G for the Micro-X Project. AJA is supported by a Department of Energy Office of Science Graduate Fellowship Program (DOE SCGF), made possible in part by  the American Recovery and Reinvestment Act of 2009, administered by ORISE-ORAU under contract no. DE-AC05-06OR23100.
DCG is supported by a NASA Space Technology Research Fellowship.
DC acknowledges support for this work provided by the Chandra GO grant GO3-14080, as well as, the National Aeronautics and Space Administration through the Smithsonian Astrophysical Observatory contract SV3-73016 to MIT for Support of the Chandra X-Ray Center, which is operated by the Smithsonian Astrophysical Observatory for and on behalf of the National Aeronautics Space Administration under contract NAS8-03060.
The XQC project is supported in part by NASA grant NNX13AH21G.

%% file: appendixXQC.tex
\appendix
\label{app:XQC}
\section{Statistical Model for XQC}
We use the method of unbinned extended maximum likelihood to fit the XQC data. The unbinned method produces equivalent results to binned $\chi^2$ fits in the limit of high statistics, but produces more reliable fits in low statistics settings where many bins would have few or zero events \citep{James2006}. The likelihood function for our background model has the form
\begin{equation}\label{eqn:XQCLikelihood}
\mathcal{L}(\textbf{S}; \{E_i\}) = \frac{e^{-\mu} \mu^N}{N!} \prod_{i=1}^N \left[ \sum_{k=1}^7 \frac{\mu_k}{\mu} P_k(\textbf{S}_k; E_i) \right],
\end{equation}
where the product is taken over the $N$ total events in the observation, and the sum is taken over each of 7 components of the background model. The values $\mu_k$ are the estimated number of events in each of the background components and $\mu \equiv \sum_k \mu_k$. The probability density functions (PDF) for each component are the $P_k(\textbf{S}_k; E_i)$, which are functions of energy and depend on the vector of parameters $\textbf{S}_k$ (with $\textbf{S} = \cup_k \textbf{S}_k$). The functional forms for each PDF are listed in Table~\ref{tab:XQCBgComponents}, and the corresponding parameters are in Table~\ref{tab:XQCBgParams}. Background PDFs which have a fixed template shape, such as the cosmic rays, do not have any parameters, so $\textbf{S}_k$ is an empty set. The Poisson term (first) is the extended likelihood term, which constrains the total expected event rate by the number of observed counts.

\begin{table*}
\begin{center}
\begin{tabular}{lll}
\hline
$k$ 		&PDF component									&Functional form \\
\hline
1		&potassium K$\alpha$, K$\beta$ cal. events in HgTe			&KDE based on pre-flight calibration\\
2		&potassium K$\alpha$ cal. events in Si substrate			&gaussian, with fitted mean and width\\
3		&potassium K$\beta$ cal. events in Si substrate			&gaussian, with fitted mean and width\\
4		&photoelectrons from cal. source escaping from absorber		&uniform from 0~keV to potassium K$\alpha$ energy\\
5		&Crab nebula 										&power law\\
6		&cosmic X-ray background (CXB) 						&power law\\
7		&cosmic rays										&spectrum derived from Geant4 simulation\\
		&												&of protons with power law above 1~GeV ($\alpha=2.7$)\\
\hline
\end{tabular}
\caption{Components of XQC background PDF in equation (\ref{eqn:XQCLikelihood}).}\label{tab:XQCBgComponents}
\end{center}
\end{table*}

The components of the background PDF are summarized in Table~\ref{tab:XQCBgComponents}, and the key model parameters are contained in Table~\ref{tab:XQCBgParams}. The calibration lines from interactions in the HgTe absorber are modeled using a gaussian kernel density estimate (KDE) based on calibration data taken in a lab after launch. An alternate model using Voigt profiles for each of the calibration lines also produces a reasonable fit, but the KDE-based model was chosen because it agrees better in the low-energy tails of the calibration lines. A similar KDE does not accurately describe the corresponding lines in the substrate, around 2.80~keV and 3.0~keV. These are more reliably modeled by gaussian PDFs with fitted means and a common fitted width. Events that interact in the substrate have suppressed energy because of charge trapping effects that depend on the neutralization state of the Si, so differences between flight and calibration data are not surprising. Lastly, the photoelectron produced by an X-ray interaction in the HgTe absorber escapes the active detector volume in about $\sim 5\%$ of events. We model this by a flat distribution extending from zero energy to the K$\alpha$ line, whose normalization is fixed to 5\% of the number of events in the K$\alpha$ and K$\beta$ absorber lines.

\begin{table*}
\begin{center}
\begin{tabular}{lll}
\hline
$k$		&Parameter											&Value \\
\hline
1		&number of K$\alpha$, K$\beta$ cal. events in HgTe				& $1281 \pm 36$ \\
2		&number of K$\alpha$ cal. events in Si substrate				& $722 \pm 29$ \\
2		&mean measured energy of K$\alpha$ cal. line in Si substrate		& $2.784 \pm 0.006$~keV \\
2 \& 3	&width of measured K$\alpha$, K$\beta$ cal. lines in Si substrate	& $0.061 \pm 0.002$~keV \\
3		&number of K$\beta$ cal. events in Si substrate				& $221 \pm 18$ \\
3		&mean measured energy of K$\beta$ cal. events in Si substrate	& $3.020 \pm 0.006$~keV \\
4		&number of cal. source events with escaping electrons in fit range	& 25 (fixed 5\% escape fraction) \\
5		&Crab spectral index \citep{Mori2004}						& 2.1 (fixed)\\
		&Crab flux at 1~keV \citep{Mori2004}						& $9.7 \pm 0.5$~photons~cm$^{-2}$~s$^{-1}$~keV$^{-1}$ (fixed) \\
		&number of Crab events									& 155 (fixed)\\
6		&CXB spectral index \citep{Hickox2006}						& 1.4 (fixed)\\
		&constraint on CXB flux at 1~keV \citep{Hickox2006}			& $10.9 \pm 1.3$~photons~cm$^{-2}$~s$^{-1}$~keV$^{-1}$~sr$^{-1}$ (fixed) \\
		&number of CXB events									& 369 (fixed) \\
7		&number of cosmic rays in fit range							& 5.2 (fixed) in 2-5~keV range\\
---		&exposure time											& 2551~pixel~s (106~s on 24~pixels) \\
\hline
\end{tabular}
\caption{Parameters of the XQC background PDF in equation (\ref{eqn:XQCLikelihood}).}\label{tab:XQCBgParams}
\end{center}
\end{table*}

We model the X-ray continuum with two power laws. One describes the flux from the Crab, using canonical parameters \citep{Mori2004}. The other describes the cosmic X-ray background, using parameters from the Chandra deep field measurement of \cite{Hickox2006}. Before fitting, the power laws must be weighted by the efficiency of both the HgTe absorber and the Si substrate. Since the events in the Si substrate appear below their true energy, this component must be shifted to lower energies by a similar fractional energy loss as the calibration lines. The resulting PDF for the continuum is given by
\begin{equation}
P_{power}(E) = \epsilon^{HgTe}(E) E^{-\alpha} + \epsilon^{subst.}(E / k) \left(\frac{E}{k} \right)^{-\alpha},
\end{equation}
where $k = E_{K\alpha}^{subst.} / 3.31$~keV is an estimate of the fractional energy loss of the K$\alpha$ calibration line. Note that this parameterization implicitly assumes that the charge trapping process in the substrate is energy-independent. Since the Crab lies $19.5^\circ$ off the observation axis, the geometrical acceptance of the detector to X-rays from the Crab is 94.2\% of the effective area for on-axis events. Because of a similar geometrical effect, the acceptance of the diffuse X-rays is 92.7\% of the effective area for on-axis events.

\begin{figure}
\begin{center}
\includegraphics[width=0.5\textwidth]{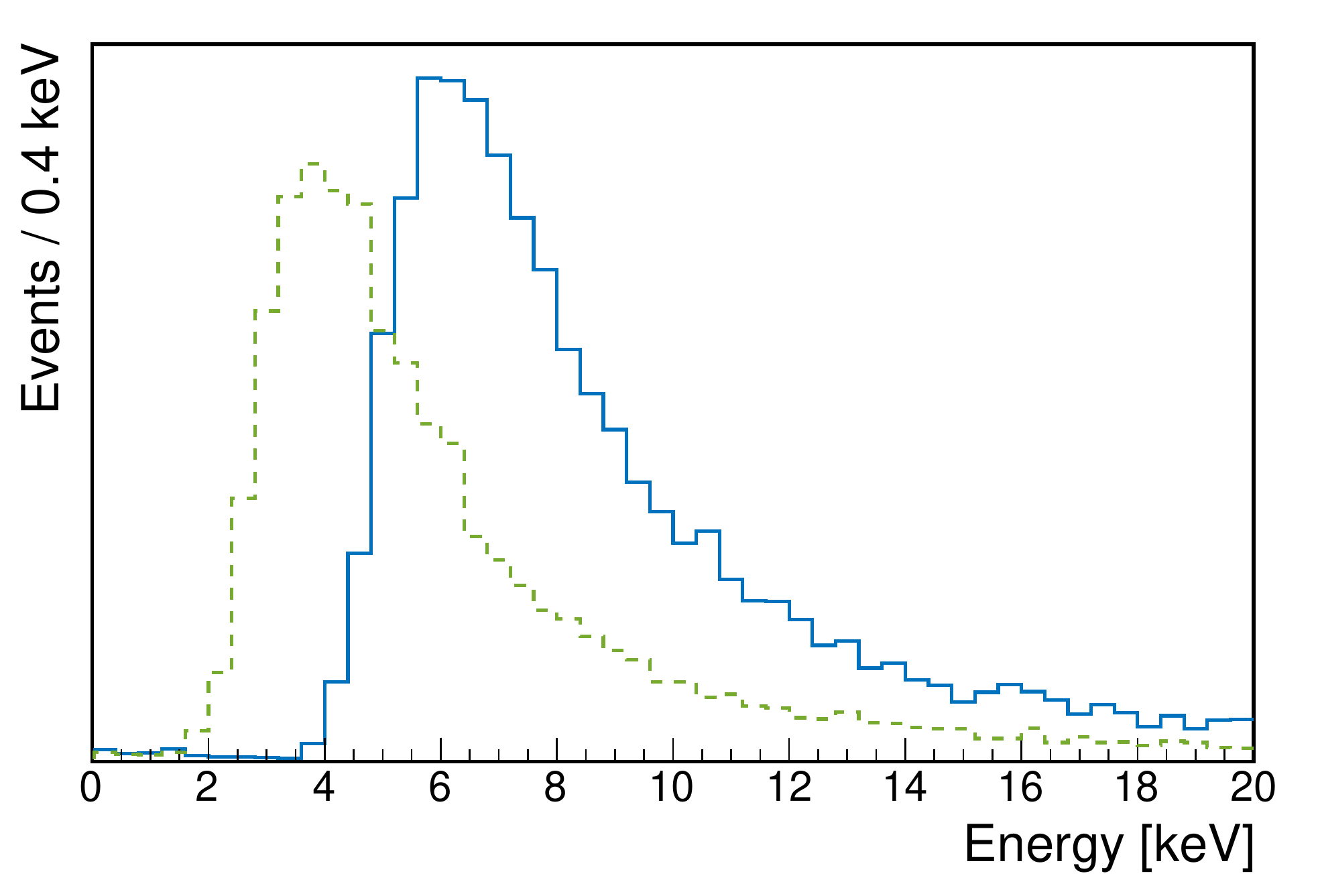}
\end{center}
\caption{Probability density function of simulated energy spectrum in the XQC \emph{(blue)} and Micro-X \emph{(dashed green)} X-ray absorbers. Geant4 \citep{Agostinelli2003} is used to simulate a power law distribution of primary cosmic ray protons \citep{Papini1996} impinging isotropically on the two absorbers. Both distributions are typical for minimum-ionizing particles. The mean energy deposited in XQC is larger than in Micro-X because of the additional 15~$\mu$m Si substrate of the HgTe absorber present in XQC but not in Micro-X. \label{fig:CosmicG4Sim}}
\end{figure}

The background component due to cosmic rays is only about 5 events in the 2-5~keV window for the full XQC exposure of 2551~pixel~s. We obtain the spectral shape of cosmic rays by simulating protons with a typical power law spectrum \citep{Papini1996} impinging on the XQC absorber and substrate. Since protons are minimum ionizing particles, this is approximately a Landau distribution with a peak around 7~keV, as shown in Figure~\ref{fig:CosmicG4Sim}.

%% file: X-ray_dm_rocket.bbl
\begin{thebibliography}{}
\expandafter\ifx\csname natexlab\endcsname\relax\def\natexlab#1{#1}\fi

\bibitem[{Abazajian {et~al.}(2001)Abazajian, Fuller, \& Tucker}]{Abazajian2001}
Abazajian, K., Fuller, G.~M., \& Tucker, W.~H. 2001, The Astrophysical Journal,
  562, 593

\bibitem[{Agostinelli {et~al.}(2003)Agostinelli, Allison, Amako, Apostolakis,
  Araujo, Arce, Asai, Axen, Banerjee, Barrand, Behner, Bellagamba, Boudreau,
  Broglia, Brunengo, Burkhardt, Chauvie, Chuma, Chytracek, Cooperman, Cosmo,
  Degtyarenko, Dell'Acqua, Depaola, Dietrich, Enami, Feliciello, Ferguson,
  Fesefeldt, Folger, Foppiano, Forti, Garelli, Giani, Giannitrapani, Gibin,
  {Gomez Cadenas}, Gonzalez, {Gracia Abril}, Greeniaus, Greiner, Grichine,
  Grossheim, Guatelli, Gumplinger, Hamatsu, Hashimoto, Hasui, Heikkinen,
  Howard, Ivanchenko, Johnson, Jones, Kallenbach, Kanaya, Kawabata, Kawabata,
  Kawaguti, Kelner, Kent, Kimura, Kodama, Kokoulin, Kossov, Kurashige, Lamanna,
  Lampen, Lara, Lefebure, Lei, Liendl, Lockman, Longo, Magni, Maire, Medernach,
  Minamimoto, {Mora de Freitas}, Morita, Murakami, Nagamatu, Nartallo,
  Nieminen, Nishimura, Ohtsubo, Okamura, O'Neale, Oohata, Paech, Perl,
  Pfeiffer, Pia, Ranjard, Rybin, Sadilov, di~Salvo, Santin, Sasaki, Savvas,
  Sawada, Scherer, Sei, Sirotenko, Smith, Starkov, Stoecker, Sulkimo, Takahata,
  Tanaka, Tcherniaev, {Safai Tehrani}, Tropeano, Truscott, Uno, Urban, Urban,
  Verderi, Walkden, Wander, Weber, Wellisch, Wenaus, Williams, Wright, Yamada,
  Yoshida, \& Zschiesche}]{Agostinelli2003}
Agostinelli, S., Allison, J., Amako, K., {et~al.} 2003, Nuclear Instruments and
  Methods in Physics Research, Section A: Accelerators, Spectrometers,
  Detectors and Associated Equipment, 506, 250

\bibitem[{{Anders} \& {Grevesse}(1989)}]{anders_1989}
{Anders}, E., \& {Grevesse}, N. 1989, \gca, 53, 197

\bibitem[{{Anderson} {et~al.}(2014){Anderson}, {Churazov}, \&
  {Bregman}}]{Anderson:2014wc}
{Anderson}, M.~E., {Churazov}, E., \& {Bregman}, J.~N. 2014, MNRAS, submitted,
  arXiv:1408.4115

\bibitem[{{Angelini} \& {White}(2003)}]{angelini_2003}
{Angelini}, L., \& {White}, N.~E. 2003, ApJL, 586, L71

\bibitem[{{Asaka} {et~al.}(2005){Asaka}, {Blanchet}, \&
  {Shaposhnikov}}]{Asaka2005}
{Asaka}, T., {Blanchet}, S., \& {Shaposhnikov}, M. 2005, Physics Letters B,
  631, 151

\bibitem[{Asaka \& Shaposhnikov(2005)}]{Asaka2005a}
Asaka, T., \& Shaposhnikov, M. 2005, Physics Letters B, 620, 17

\bibitem[{Barlow(1990)}]{Barlow1990}
Barlow, R. 1990, Nuclear Instruments and Methods in Physics Research A, 297,
  496

\bibitem[{Bearden(1967)}]{bearden_1967}
Bearden, J.~A. 1967, Rev. Mod. Phys., 39, 78

\bibitem[{Berger {et~al.}(2010)Berger, Hubbell, Seltzer, Chang, Coursey,
  Sukumar, Zucker, \& Olsen}]{berger:10:xcom}
Berger, M.~J., Hubbell, J.~H., Seltzer, S.~M., {et~al.} 2010, {XCOM: Photon
  Cross Sections Database} (National Institute of Standards and Technology)

\bibitem[{{Berlin} {et~al.}(2015){Berlin}, {DiFranzo}, \&
  {Hooper}}]{Berlin2015}
{Berlin}, A., {DiFranzo}, A., \& {Hooper}, D. 2015, \prd, 91, 075018

\bibitem[{Bernal \& Palomares-Ruiz(2012)}]{Bernal2012}
Bernal, N., \& Palomares-Ruiz, S. 2012, Journal of Cosmology and Astroparticle
  Physics, 2012, 006

\bibitem[{{Boyarsky} {et~al.}(2007){Boyarsky}, {den Herder}, {Neronov}, \&
  {Ruchayskiy}}]{Boyarsky2006a}
{Boyarsky}, A., {den Herder}, J.-W., {Neronov}, A., \& {Ruchayskiy}, O. 2007,
  Astroparticle Physics, 28, 303

\bibitem[{Boyarsky {et~al.}(2014{\natexlab{a}})Boyarsky, Franse, Iakubovskyi,
  \& Ruchayskiy}]{Boyarsky2014a}
Boyarsky, A., Franse, J., Iakubovskyi, D., \& Ruchayskiy, O.
  2014{\natexlab{a}}, arXiv:1408.2503

\bibitem[{Boyarsky {et~al.}(2006)Boyarsky, Neronov, Ruchayskiy, Shaposhnikov,
  \& Tkachev}]{Boyarsky2006}
Boyarsky, A., Neronov, A., Ruchayskiy, O., Shaposhnikov, M., \& Tkachev, I.
  2006, Physical Review Letters, 97, 261302

\bibitem[{Boyarsky {et~al.}(2014{\natexlab{b}})Boyarsky, Ruchayskiy,
  Iakubovskyi, \& Franse}]{Boyarsky2014}
Boyarsky, A., Ruchayskiy, O., Iakubovskyi, D., \& Franse, J.
  2014{\natexlab{b}}, Phys. Rev. Lett., 113, 251301

\bibitem[{Boyarsky {et~al.}(2009)Boyarsky, Ruchayskiy, \&
  Shaposhnikov}]{Boyarsky2009}
Boyarsky, A., Ruchayskiy, O., \& Shaposhnikov, M. 2009, Annual Review of
  Nuclear and Particle Science, 59, 191

\bibitem[{Bulbul {et~al.}(2014{\natexlab{a}})Bulbul, Markevitch, Foster, Smith,
  Loewenstein, \& Randall}]{Bulbul2014}
Bulbul, E., Markevitch, M., Foster, A., {et~al.} 2014{\natexlab{a}}, The
  Astrophysical Journal, 789, 13

\bibitem[{Bulbul {et~al.}(2014{\natexlab{b}})Bulbul, Markevitch, Foster, Smith,
  Loewenstein, \& Randall}]{Bulbul:2014wj}
Bulbul, E., Markevitch, M., Foster, A.~R., {et~al.} 2014{\natexlab{b}},
  arXiv:1409.4143

\bibitem[{{Cackett} {et~al.}(2009){Cackett}, {Miller}, {Homan}, {van der Klis},
  {Lewin}, {M{\'e}ndez}, {Raymond}, {Steeghs}, \& {Wijnands}}]{cackett_2009}
{Cackett}, E.~M., {Miller}, J.~M., {Homan}, J., {et~al.} 2009, \apj, 690, 1847

\bibitem[{{Carlson} {et~al.}(2015){Carlson}, {Hooper}, \&
  {Linden}}]{Carlson2014}
{Carlson}, E., {Hooper}, D., \& {Linden}, T. 2015, \prd, 91, 061302

\bibitem[{Carlson {et~al.}(2015)Carlson, Jeltema, \& Profumo}]{Carlson:2014vl}
Carlson, E., Jeltema, T., \& Profumo, S. 2015, Journal of Cosmology and
  Astroparticle Physics, 2015, 009

\bibitem[{Cholis {et~al.}(2014)Cholis, Hooper, \& Linden}]{Cholis2014}
Cholis, I., Hooper, D., \& Linden, T. 2014, arXiv:1407.5625

\bibitem[{Christian \& Swank(1997)}]{ScoX-2}
Christian, D., \& Swank, J. 1997, The Astrophysical Journal Supplement Series,
  109, 177

\bibitem[{{Costantini} {et~al.}(2012){Costantini}, {Pinto}, {Kaastra}, {in't
  Zand}, {Freyberg}, {Kuiper}, {M{\'e}ndez}, {de Vries}, \&
  {Waters}}]{costantini_2012}
{Costantini}, E., {Pinto}, C., {Kaastra}, J.~S., {et~al.} 2012, \aap, 539, A32

\bibitem[{Cowan {et~al.}(2011)Cowan, Cranmer, Gross, \& Vitells}]{Cowan2011}
Cowan, G., Cranmer, K., Gross, E., \& Vitells, O. 2011, The European Physical
  Journal C, 71, 1

\bibitem[{{Crowder} {et~al.}(2012){Crowder}, {Barger}, {Brandl}, {Eckart},
  {Galeazzi}, {Kelley}, {Kilbourne}, {McCammon}, {Pfendner}, {Porter}, {Rocks},
  {Szymkowiak}, \& {Teplin}}]{Crowder2012}
{Crowder}, S.~G., {Barger}, K.~A., {Brandl}, D.~E., {et~al.} 2012, \apj, 758,
  143

\bibitem[{Dickey \& Lockman(1990)}]{XRayHandbook}
Dickey, J.~M., \& Lockman, F.~J. 1990, Annual Review of Astronomy and
  Astrophysics, 28, 215

\bibitem[{Dodelson \& Widrow(1994)}]{Dodelson1994}
Dodelson, S., \& Widrow, L.~M. 1994, Physical Review Letters, 72, 17

\bibitem[{Egron {et~al.}(2013)Egron, Salvo, Motta, Burderi, Papitto, Duro,
  D'A\`{i}, Riggio, Belloni, Iaria, Robba, Piraino, \& Santangelo}]{4U1705-44}
Egron, E., Salvo, T.~D., Motta, S., {et~al.} 2013, Astronomy and Astrophysics,
  550

\bibitem[{Ercan(1988)}]{GX17+2}
Ercan, E. 1988, Astrophysics and Space Science, 147, 145

\bibitem[{Feldman \& Cousins(1998)}]{Feldman1998}
Feldman, G.~J., \& Cousins, R.~D. 1998, Physical Review D, 57, 3873

\bibitem[{Finkbeiner \& Weiner(2007)}]{Finkbeiner2007}
Finkbeiner, D.~P., \& Weiner, N. 2007, Physical Review D, 76, 083519

\bibitem[{Finkbeiner \& Weiner(2014)}]{Finkbeiner2014}
---. 2014, arXiv:1402.6671

\bibitem[{{Foster} {et~al.}(2012){Foster}, {Ji}, {Smith}, \&
  {Brickhouse}}]{foster_2012}
{Foster}, A.~R., {Ji}, L., {Smith}, R.~K., \& {Brickhouse}, N.~S. 2012, \apj,
  756, 128

\bibitem[{{Heine} {et~al.}(2014){Heine}, {Figueroa-Feliciano}, {Rutherford},
  {Wikus}, {Oakley}, {Porter}, \& {McCammon}}]{Heine2014}
{Heine}, S.~N.~T., {Figueroa-Feliciano}, E., {Rutherford}, J.~M., {et~al.}
  2014, Journal of Low Temperature Physics, 176, 1082

\bibitem[{Hickox \& Markevitch(2006)}]{Hickox2006}
Hickox, R.~C., \& Markevitch, M. 2006, The Astrophysical Journal, 645, 95

\bibitem[{Horiuchi {et~al.}(2014)Horiuchi, Humphrey, O\~{n}orbe, Abazajian,
  Kaplinghat, \& Garrison-Kimmel}]{Horiuchi2014}
Horiuchi, S., Humphrey, P.~J., O\~{n}orbe, J., {et~al.} 2014, Physical Review
  D, 89, 025017

\bibitem[{{Iaria} {et~al.}(2005){Iaria}, {di Salvo}, {Robba}, {Lavagetto},
  {Burderi}, {Stella}, \& {van der Klis}}]{iaria_2005}
{Iaria}, R., {di Salvo}, T., {Robba}, N.~R., {et~al.} 2005, \aap, 439, 575

\bibitem[{James(2006)}]{James2006}
James, F. 2006, {Statistical Methods in Experimental Physics}, 2nd edn.
  (Singapore: World Scientific)

\bibitem[{Jeltema \& Profumo(2014)}]{Jeltema:2014wr}
Jeltema, T., \& Profumo, S. 2014, arXiv:arXiv:1411.1759v1

\bibitem[{{Kaastra} \& {Mewe}(1993)}]{kaastra_1993}
{Kaastra}, J.~S., \& {Mewe}, R. 1993, \aaps, 97, 443

\bibitem[{{Kitayama} {et~al.}(2014){Kitayama}, {Bautz}, {Markevitch},
  {Matsushita}, {Allen}, {Kawaharada}, {McNamara}, {Ota}, {Akamatsu}, {de
  Plaa}, {Galeazzi}, {Madejski}, {Main}, {Miller}, {Nakazawa}, {Russell},
  {Sato}, {Sekiya}, {Simionescu}, {Tamura}, {Uchida}, {Ursino}, {Werner},
  {Zhuravleva}, {ZuHone}, \& {on behalf of the ASTRO-H Science Working
  Group}}]{Kitayama2014}
{Kitayama}, T., {Bautz}, M., {Markevitch}, M., {et~al.} 2014, ArXiv e-prints,
  arXiv:1412.1176

\bibitem[{Kong {et~al.}(2006)Kong, Charles, L.Homer, Kuulkers, \&
  O'Donoghue}]{GX9+9}
Kong, A., Charles, P., L.Homer, Kuulkers, E., \& O'Donoghue, D. 2006, Monthly
  Notices of the Royal Astronomical Society, 368, 781

\bibitem[{{Kushino} {et~al.}(2002){Kushino}, {Ishisaki}, {Morita}, {Yamasaki},
  {Ishida}, {Ohashi}, \& {Ueda}}]{kushino_2002}
{Kushino}, A., {Ishisaki}, Y., {Morita}, U., {et~al.} 2002, \pasj, 54, 327

\bibitem[{Lovell {et~al.}(2015)Lovell, Bertone, Boyarsky, Jenkins, \&
  Ruchayskiy}]{Lovell:2014tx}
Lovell, M.~R., Bertone, G., Boyarsky, A., Jenkins, A., \& Ruchayskiy, O. 2015,
  Monthly Notices of the Royal Astronomical Society, 451, 1573

\bibitem[{Mainardi {et~al.}(2010)Mainardi, Paizis, Farinelli, Kuulkers,
  Rodriguez, Hannikainen, Savolainen, Piraino, Bazzano, \& Santangelo}]{GX9+1}
Mainardi, L.~I., Paizis, A., Farinelli, R., {et~al.} 2010, Astronomy and
  Astrophysics, 512

\bibitem[{Malyshev {et~al.}(2014)Malyshev, Neronov, \&
  Eckert}]{Malyshev:2014vf}
Malyshev, D., Neronov, A., \& Eckert, D. 2014, Physical Review D, 103506, 1

\bibitem[{McCammon {et~al.}(2002)McCammon, Almy, Apodaca, Bergmann~Tiest, Cui,
  Deiker, Galeazzi, Juda, Lesser, Mihara, Morgenthaler, Sanders, Zhang,
  Figueroa-Feliciano, Kelley, Moseley, Mushotzky, Porter, Stahle, \&
  Szymkowiak}]{McCammon:2002p160}
McCammon, D., Almy, R., Apodaca, E., {et~al.} 2002, The Astrophysical Journal,
  576, 188

\bibitem[{{Mihara} {et~al.}(2014){Mihara}, {Sugizaki}, {Matsuoka}, {Tomida},
  {Ueno}, {Negoro}, {Yoshida}, {Tsunemi}, {Nakajima}, {Ueda}, \&
  {Yamauchi}}]{Mihara2014}
{Mihara}, T., {Sugizaki}, M., {Matsuoka}, M., {et~al.} 2014, in Society of
  Photo-Optical Instrumentation Engineers (SPIE) Conference Series, Vol. 9144,
  Society of Photo-Optical Instrumentation Engineers (SPIE) Conference Series,
  1

\bibitem[{Moneta {et~al.}(2010)Moneta, Belasco, Cranmer, Kreiss, Lazzaro,
  Piparo, Schott, Verkerke, \& Wolf}]{Moneta2011new}
Moneta, L., Belasco, K., Cranmer, K., {et~al.} 2010, in Proceedings of the 13th
  International Workshop on Advanced Computing and Analysis Techniques in
  Physics (Proceedings of Science), 57

\bibitem[{Moneta {et~al.}(2011)Moneta, Belasco, Cranmer, Kreiss, Lazzaro,
  Piparo, Schott, Verkerke, \& Wolf}]{Moneta2011}
Moneta, L., Belasco, K., Cranmer, K., {et~al.} 2011, arXiv:1009.1003v2

\bibitem[{Mori {et~al.}(2004)Mori, Burrows, Hester, Pavlov, Shibata, \&
  Tsunemi}]{Mori2004}
Mori, K., Burrows, D.~N., Hester, J.~J., {et~al.} 2004, The Astrophysical
  Journal, 609, 186

\bibitem[{{Narita} {et~al.}(2001){Narita}, {Grindlay}, \&
  {Barret}}]{narita_2001}
{Narita}, T., {Grindlay}, J.~E., \& {Barret}, D. 2001, \apj, 547, 420

\bibitem[{Nesti \& Salucci(2013)}]{Nesti2013}
Nesti, F., \& Salucci, P. 2013, Journal of Cosmology and Astroparticle Physics,
  7, 16

\bibitem[{{Ng} {et~al.}(2010){Ng}, {D{\'{\i}}az Trigo}, {Cadolle Bel}, \&
  {Migliari}}]{ng_2010}
{Ng}, C., {D{\'{\i}}az Trigo}, M., {Cadolle Bel}, M., \& {Migliari}, S. 2010,
  \aap, 522, A96

\bibitem[{{Ng} {et~al.}(2015){Ng}, {Horiuchi}, {Gaskins}, {Smith}, \&
  {Preece}}]{Ng2015}
{Ng}, K.~C.~Y., {Horiuchi}, S., {Gaskins}, J.~M., {Smith}, M., \& {Preece}, R.
  2015, ArXiv e-prints, arXiv:1504.04027

\bibitem[{{Nobukawa} {et~al.}(2010){Nobukawa}, {Koyama}, {Tsuru}, {Ryu}, \&
  {Tatischeff}}]{nobukawa_2010}
{Nobukawa}, M., {Koyama}, K., {Tsuru}, T.~G., {Ryu}, S.~G., \& {Tatischeff}, V.
  2010, \pasj, 62, 423

\bibitem[{Pal \& Wolfenstein(1982)}]{Pal1982}
Pal, P.~B., \& Wolfenstein, L. 1982, Physical Review D, 25, 766

\bibitem[{Papini {et~al.}(1996)Papini, Grimani, \& Stephens}]{Papini1996}
Papini, P., Grimani, C., \& Stephens, S. 1996, Il Nuovo Cimento C, 19, 367

\bibitem[{{Piraino} {et~al.}(2012){Piraino}, {Santangelo}, {Kaaret},
  {M{\"u}ck}, {D'A{\`i}}, {Di Salvo}, {Iaria}, {Robba}, {Burderi}, \&
  {Egron}}]{piraino_2012}
{Piraino}, S., {Santangelo}, A., {Kaaret}, P., {et~al.} 2012, \aap, 542, L27

\bibitem[{Piraino {et~al.}(2012)Piraino, Santangelo, Kaaret, M{\"u}ck,
  D'A\`{i}, Salvo, Iaria, Robba, Burderi, \& Egron}]{XSgrX1}
Piraino, S., Santangelo, A., Kaaret, P., {et~al.} 2012, Astronomy and
  Astrophysics, 542

\bibitem[{Read(2014)}]{Read2014}
Read, J.~I. 2014, Journal of Physics G: Nuclear and Particle Physics, 41,
  063101

\bibitem[{Remillard \& McClintock(2006)}]{LXRBSpec}
Remillard, R.~A., \& McClintock, J.~E. 2006, Annual Review of Astrophysics, 44,
  49

\bibitem[{Riemer-Sorensen(2014)}]{RiemerSorensen:2014us}
Riemer-Sorensen, S. 2014, arXiv:1405.7943

\bibitem[{Seon {et~al.}(1997)Seon, Min, Yoshida, Makino, Lewin, van~der Klis,
  \& Paradijs}]{3A1735-444}
Seon, K.-I., Min, K.-W., Yoshida, K., {et~al.} 1997, The Astrophysical Journal,
  479, 398

\bibitem[{Seon {et~al.}(1995)Seon, Min, Yoshida, Makino, van~der Klis,
  Paradijs, \& Lewin}]{3A1755-338}
---. 1995, The Astrophysical Journal, 454, 463

\bibitem[{Shi \& Fuller(1999)}]{Shi1999}
Shi, X., \& Fuller, G. 1999, Physical Review Letters, 82, 2832

\bibitem[{{Smith} {et~al.}(2001){Smith}, {Brickhouse}, {Liedahl}, \&
  {Raymond}}]{smith_2001}
{Smith}, R.~K., {Brickhouse}, N.~S., {Liedahl}, D.~A., \& {Raymond}, J.~C.
  2001, ApJL, 556, L91

\bibitem[{Sriram {et~al.}(2012)Sriram, Choi, \& Rao}]{GX5-1}
Sriram, K., Choi, C., \& Rao, A. 2012, The Astrophysical Journal Supplement
  Series, 200

\bibitem[{Strigari(2013)}]{Strigari2013}
Strigari, L.~E. 2013, Physics Reports, 531, 1

\bibitem[{{Takahashi} {et~al.}(2014){Takahashi}, {Mitsuda}, {Kelley},
  {Aharonian}, {Akamatsu}, {Akimoto}, {Allen}, {Anabuki}, {Angelini}, {Arnaud},
  \& et~al.}]{Takahashi2014}
{Takahashi}, T., {Mitsuda}, K., {Kelley}, R., {et~al.} 2014, in Society of
  Photo-Optical Instrumentation Engineers (SPIE) Conference Series, Vol. 9144,
  Society of Photo-Optical Instrumentation Engineers (SPIE) Conference Series,
  25

\bibitem[{Thompson \& Vaughan(2001)}]{xdb2001}
Thompson, A.~C., \& Vaughan, D., eds. 2001, {X-ray Data Booklet}, 2nd edn.
  (Lawrence Berkeley National Laboratory, University of California)

\bibitem[{{Uchiyama} {et~al.}(2013){Uchiyama}, {Nobukawa}, {Tsuru}, \&
  {Koyama}}]{uchiyama_2013}
{Uchiyama}, H., {Nobukawa}, M., {Tsuru}, T.~G., \& {Koyama}, K. 2013, \pasj,
  65, 19

\bibitem[{{Ueda} {et~al.}(2005){Ueda}, {Mitsuda}, {Murakami}, \&
  {Matsushita}}]{ueda_2005}
{Ueda}, Y., {Mitsuda}, K., {Murakami}, H., \& {Matsushita}, K. 2005, \apj, 620,
  274

\bibitem[{{Urban} {et~al.}(2014){Urban}, {Werner}, {Allen}, {Simionescu},
  {Kaastra}, \& {Strigari}}]{Urban:2014va}
{Urban}, O., {Werner}, N., {Allen}, S.~W., {et~al.} 2014, MNRAS, submitted,
  arXiv:1411.0050

\bibitem[{Verkerke \& Kirkby(2003)}]{Verkerke2003}
Verkerke, W., \& Kirkby, D. 2003, arXiv:0306116

\bibitem[{{Voges} {et~al.}(1999){Voges}, {Aschenbach}, {Boller},
  {Br{\"a}uninger}, {Briel}, {Burkert}, {Dennerl}, {Englhauser}, {Gruber},
  {Haberl}, {Hartner}, {Hasinger}, {K{\"u}rster}, {Pfeffermann}, {Pietsch},
  {Predehl}, {Rosso}, {Schmitt}, {Tr{\"u}mper}, \& {Zimmermann}}]{voges_1999}
{Voges}, W., {Aschenbach}, B., {Boller}, T., {et~al.} 1999, \aap, 349, 389

\bibitem[{Wijnands {et~al.}(2001)Wijnands, Miller, Markwardt, Lewin, \& van~der
  Klis}]{KS1731-260}
Wijnands, R., Miller, J.~M., Markwardt, C., Lewin, W.~H., \& van~der Klis, M.
  2001, The Astrophysical Journal Letters, 560, 159

\bibitem[{{Wilms} {et~al.}(2000){Wilms}, {Allen}, \& {McCray}}]{wilms_2000}
{Wilms}, J., {Allen}, A., \& {McCray}, R. 2000, \apj, 542, 914

\bibitem[{{Yamauchi} \& {Nakamura}(2004)}]{yamauchi_2004}
{Yamauchi}, S., \& {Nakamura}, E. 2004, \pasj, 56, 803

\bibitem[{Yamauchi \& Nakamura(2004)}]{G1734-275}
Yamauchi, S., \& Nakamura, E. 2004, Publications of the Astronomical Society of
  Japan, 56, 803

\end{thebibliography}
